\documentclass[journal]{IEEEtran}
\ifCLASSINFOpdf
   \usepackage[pdftex]{graphicx}
   \DeclareGraphicsExtensions{.pdf,.jpeg,.png}
\else
  \usepackage[dvips]{graphicx}
  \DeclareGraphicsExtensions{.eps}
\fi

\usepackage{pbox}
\usepackage{booktabs}
\usepackage{tabularx}
\usepackage{blindtext, subfig}
\usepackage[dvipsnames]{xcolor}
\usepackage{url}
\usepackage[strings]{underscore}
\usepackage{amsmath}
\usepackage{multirow}
\usepackage{colortbl}

\usepackage{xcolor}
\colorlet{ColorVariable1}{black}
\colorlet{ColorVariable2}{black}
\colorlet{ColorVariable3}{black}

\usepackage{array}
\usepackage{graphicx}
\graphicspath{ {./imgs/} }
\usepackage{caption}
\begin{document}
\title{Multichannel Speech Enhancement by Raw Waveform-mapping using Fully Convolutional Networks}

\author{\IEEEauthorblockN{Chang-Le Liu\IEEEauthorrefmark{1},
Sze-Wei Fu\IEEEauthorrefmark{2},
You-Jin Li\IEEEauthorrefmark{2}, 
Jen-Wei Huang\IEEEauthorrefmark{3},
Hsin-Min Wang\IEEEauthorrefmark{4},
Yu Tsao\IEEEauthorrefmark{2},}\par
\IEEEauthorblockA{\IEEEauthorrefmark{1}Department of Electrical Engineering, National Taiwan University, Taipei, Taiwan}\\
\IEEEauthorblockA{\IEEEauthorrefmark{2}Research Center for Information Technology Innovation at Academia Sinica, Taipei, Taiwan}\\
\IEEEauthorblockA{\IEEEauthorrefmark{3}Department of Electrical Engineering, National Cheng-Kung University, Tainan, Taiwan}\\ 
\IEEEauthorblockA{\IEEEauthorrefmark{4}Institute of Information Science, Academia Sinica, Taipei, Taiwan}% <-this % stops an unwanted space
\thanks{Manuscript received \today; revised \today.}}

\markboth{Journal of \LaTeX\ Class Files,~Vol.~14, No.~8, May~2016}%
{Shell \MakeLowercase{\textit{et al.}}: Bare Demo of IEEEtran.cls for IEEE Journals}

\maketitle

\begin{abstract}
In recent years, waveform-mapping-based speech enhancement (SE) methods have garnered significant attention. These methods generally use a deep learning model to directly process and reconstruct speech waveforms. Because both the input and output are in waveform format, the waveform-mapping-based SE methods can overcome the distortion caused by imperfect phase estimation, which may be encountered in spectral-mapping-based SE systems. So far, most waveform-mapping-based SE methods have focused on single-channel tasks. In this paper, we propose a novel fully convolutional network (FCN) with Sinc and dilated convolutional layers (termed SDFCN) for multichannel SE that operates in the time domain. We also propose an extended version of SDFCN, called the residual SDFCN (termed rSDFCN). The proposed methods are evaluated on three multichannel SE tasks, namely the dual-channel inner-ear microphones SE task, the distributed microphones SE task, \textcolor{ColorVariable1}{and the CHiME-3 dataset.} The experimental results confirm the outstanding denoising capability of the proposed SE systems on the three tasks and the benefits of using the residual architecture on the overall SE performance.
\end{abstract}

% Note that keywords are not normally used for peerreview papers.

% For peer review papers, you can put extra information on the cover
% page as needed:
\ifCLASSOPTIONpeerreview
\begin{center} \bfseries EDICS Category: 3-BBND \end{center}
\fi
%
% For peerreview papers, this IEEEtran command inserts a page break and
% creates the second title. It will be ignored for other modes.
\IEEEpeerreviewmaketitle

\section{Introduction}
\IEEEPARstart{S}{peech} -related applications for both human-human and human-machine interfaces have garnered significant attention in recent years. However, speech signals are easily distorted by additive or convolutional noises or recording devices, and such distortion constrains the achievable performance of these applications. To address this issue, numerous speech enhancement (SE) algorithms have been derived to improve the quality and intelligibility of distorted speech and are widely used as a preprocessor in speech-related applications, such as speech coding \cite{Zhao2018Convolutional}, \cite{Li2011Comparative}, assistive hearing devices \cite{Chen2016Large-scale}, \cite{Lai2017Deep}, and automatic speech recognition (ASR) \cite{Li2015Robust}. Generally speaking, SE methods can be divided into two categories. The first category adopts a single channel (also termed monaural) while the second category uses multiple microphones (also termed multichannel) to perform SE. \par
Traditional single-channel-based SE methods were derived based on the characteristics and statistical assumptions of clean speech and noise signals. Well-known approaches include spectral-subtraction \cite{Boll1979Suppression}, the Wiener filter \cite{Krishnamoorthi2010auditory-domain}, \cite{McAulay1980Speech}, and the minimum mean square error (MMSE) \cite{Ephraim1984Speech}. Another category of successful SE approaches is subspace-based methods, which aim to separate noisy speech into two subspaces, one for clean speech and the other for noise components. The clean speech is then restored based on the information in the clean-speech subspace. Notable subspace techniques include generalized subspace approaches with prewhitening \cite{Loizou2003generalized}, the Karhunen-Loeve transform \cite{Rezayee2001adaptive}, and principal component analysis (PCA) \cite{Vetter1999Single}. \par
In recent years, machine-learning-based algorithms have been popularly used in the SE field. Unlike traditional methods, a machine-learning-based SE approach generally prepares a denoising model in a data-driven manner without imposing strong statistical constraints. Well-known machine-learning-based models include non-negative matrix factorization \cite{Mohammadiha2013Supervised}, compressive sensing \cite{Wang2016Compressive}, sparse coding \cite{Sigg2010Speech}, and robust principal component analysis (RPCA) \cite{Huang2012Singing-voice}. More recently, deep learning models have been applied to the SE field. Owing to their outstanding nonlinear mapping capability, deep-learning-based SE methods have demonstrated notable performance improvements over traditional statistical methods and other machine-learning-based methods. Well-known deep-learning-based models include the deep de-noising autoencoder (DDAE) \cite{Lu2014Ensemble}, \cite{Lu2013Speech}, deep fully connected networks \cite{Kolbak2017Speech,LiuExperiments,Xu2014Experimental,Xu2015Regression}, recurrent neural networks \cite{Campolucci1999On-line}, \cite{Weninger2014Single-channel}, convolutional neural networks \cite{Fu2017Complex}, \cite{Fu2016SNR-Aware}, and long short-term memory \cite{Eyben2013Real-life,Weninger2015Speech,Chen2015Speech,Sun2017Multiple-target}. \par
Different from single-channel SE methods, the multichannel ones utilize information from plural channels to enhance the target speech signal. Among the multichannel SE methods, beamforming \cite{Bitzer1999Multi-Microphone, Liu1995Room, Hoshuyama1999robust} is a popular method that exploits spatial information from multiple microphones to attenuate inference and noise signals. In addition to beamforming, other effective methods are based on a coherence algorithm that calculates the correlation of two input signals to estimate a filter to attenuate the interference components \cite{Yousefian2011Dual-Microphone,Kailath1985Adaptive}. Meanwhile, Li et al. proposed a method of using distributed-microphones for in-vehicle SE [36]. They argued the clean speech signals acquired by distributed-microphones are similar to each other while the noise signals acquired by distributed-microphones are irrelevant to each other. Therefore, the RPCA algorithm \cite{Huang2012Singing-voice} is applied to the matrix formed by the acquired noisy signals from multiple channels to separate clean speech and noise components \cite{Li2018Distributed-microphones}.\par
More recently, deep learning-based models also exhibit encouraging performance in multichannel SE tasks. Araki et al. showed that multichannel audio features can effectively improve the performance of the denoising auto-encoder (DAE) \cite{Vincent2008Extracting} based SE approach \cite{Araki2015Exploring}. Wang and Wang proposed a deep learning-based time-frequency (T-F) masking SE method that estimates robust time delay of arrival over multiple singly-enhanced speech signals to obtain directional features and hence the beam-formed signals. The enhancement is carried out by combining spectral and directional features \cite{Wang2018All-Neural}. Although the above-mentioned multichannel SE approaches have been able to provide satisfactory performance, they are performed in the frequency domain, i.e., they typically use the phase from the noisy input and require additional processing to convert the speech waveform into spectral features. To avoid imperfect phase estimation and reduce online processing, waveform-mapping-based audio signal processing methods have been developed. For example, in \cite{Fu2017Complex,Fu2018End-to-End,Pascual2017SEGAN:,Rethage2017Wavenet, Qian2017Speech}, a fully convolutional network (FCN) model was used to enhance on the noisy waveform to generate an enhanced waveform, and in \cite{Grais2017Multi-Resolution,Grais2018Raw}, the FCN model was used to separate a singing voice from mono or stereo music. \par
In the present work, we propose a novel fully convolutional network that incorporates Sinc convolutional filters (termed SincConv) and dilated convolutional filters, to perform multichannel SE in the time domain. Therefore, the model is called Sinc dilated FCN (termed SDFCN). In addition, we derive an extended system from the SDFCN system. The extended system structures a residual architecture in which SDFCN is used to estimate and compensate for the residual components of the enhanced speech from a primary SE model. Therefore, it is named residual SDFCN (termed rSDFCN). We evaluate the proposed models on three multichannel SE tasks: inner-ear microphones (termed the IEM-SE task), distributed-microphones (termed the DM-SE task)\textcolor{ColorVariable1}{, and the CHiME-3 dataset \cite{barker2015third}}. For these tasks, the proposed SE models take inputs from multiple channels to generate a single-channel waveform with higher quality and intelligibility than individual noisy inputs. Two standardized metrics are used in the evaluation: short-time objective intelligibility (STOI) \cite{Taal2010short-time,Taal2011Algorithm} and perceptual estimation of speech quality (PESQ) \cite{RixPerceptual}. In addition, we conduct subjective listening and speech recognition tests with the enhanced speech signals. Our experimental results confirm the outstanding denoising capability of the proposed SDFCN and rSDFCN models in all three multichannel SE tasks, demonstrating the benefits of using the residual architecture on the overall SE performance. 

The remainder of this paper is organized as follows. Section 2 reviews the related works. Section 3 presents the concept and architectures of the proposed SDFCN and rSDFCN models. Section 4 presents the experimental setup and results. Finally, Section 5 concludes this work.

\section{RELATED WORKS}
Given a clean speech signal $\mathbf{x}$, the degraded signal can be formulated as $\mathbf{y}=g(\mathbf{x})$, where $g$ denotes the degradation function. The goal of SE is to find a function that maps y to $\mathbf{\hat{x}}$ that approximates $\mathbf{x}$ as close as possible. In this section, we review related works, including the FCN-based waveform-mapping-based SE method, SincConv filters, and dilated convolutional filters.  
\subsection{Waveform-mapping-based SE}
Previous studies have shown that the FCN model is suitable for waveform-mapping-based SE because the convolutional layers can more effectively characterize the local information of neighboring input regions \cite{Fu2017Raw}. FCN is a modified convolutional neural network (CNN) model in which the fully connected layers in CNN are completely replaced by the convolutional layers, as shown in Fig. 1. In FCN, the relation between  each sample point \textcolor{ColorVariable1}{ $\mathbf{\hat{x}}_t$ of the output $\mathbf{\hat{x}}$ and the last connected hidden nodes $\mathbf{h_t}\in R^{L\times1}$ can be represented by
\begin{equation}
	\mathbf{\hat{x}}_t= \mathbf{v}^T \mathbf{h}_t + b,
\end{equation}
where $\mathbf{v}\in R^{L\times1}$ denotes a convolutional filter, $b$ is a bias term, and $L$ is the size of the filter. Note that $\mathbf{v}$ and $b$ are} shared in the convolution operation and are fixed for every output. Because the pooling step may reduce the precision of speech signal reconstruction, we did not apply any pooling operations (e.g., WaveNet \cite{Oord2016WaveNet:}) to perform SE when using FCN. For more details about the structure of the FCN model applied to waveform-mapping-based SE, please refer to previous works \cite{Fu2017Raw,Fu2018End-to-End,Oord2016WaveNet:}.
% for two-column fig:
% https://tex.stackexchange.com/questions/30985/displaying-a-wide-figure-in-a-two-column-document

\begin{figure}
  \centering
  \includegraphics[width=0.5\textwidth]{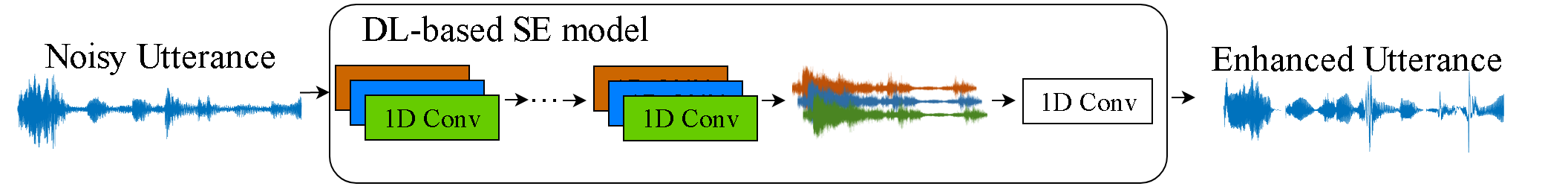}
  \caption{A waveform-mapping-based SE system.}
\end{figure}

\subsection{SincConv Filters}
As mentioned above, convolutional filters are often used to process raw-waveforms. When the CNN model is too deep or the training data is insufficient, the filters of the first few layers may not be well learned because of the vanishing gradient issue. \textcolor{ColorVariable2}{To overcome this issue, Ravanelli et al. \cite{Ravanelli2018Speaker} recently proposed a novel convolutional architecture, called SincNet.} Unlike conventional CNN models that learn all filters based on training data, SincNet predefines the filters of the first few layers to model the rectangular band-pass filter-banks in the frequency domain. \textcolor{ColorVariable1}{Specifically, assuming that the filter function of the first layer is $\mathbf{v}$, which will be convolved with the input signal $\mathbf{y}$, then $\mathbf{v}$ can be written as follows:}

\begin{align*}
    \mathbf{v}&=\mathbf{s}\circ \mathbf{w}\\
	\mathbf{s}_t&=2f_{low} \mathrm{sinc}(2\pi f_{low} t)-2f_{high} \mathrm{sinc}(2\pi f_{high} t)\\
	\mathbf{w}_t&=0.54-0.46\mathrm{cos}(\frac{\pi t}{L})
\end{align*}

\noindent\textcolor{ColorVariable1}{where $\circ$ is component-wise multiplication, $L$ is the filter length}, and $f_{low}$ and $f_{high}$ are the low and high cutoff frequencies learned during training, respectively. Obviously, this architecture is much more efficient because each filter in the first layer only consists of two coefficients rather than $L$ (the original filter length) coefficients. In [51], it was shown that SincNet converged faster in training and performed better in testing than CNN on a speaker recognition task when the input was raw speech waveform.The smaller number of neurons enables SincNet to be well trained even on a limited training dataset \cite{Ravanelli2018Speaker}.

\subsection{Dilated Convolution}

Previous works, such as WaveNet \cite{Oord2016WaveNet:}, Conv-TasNet \cite{Luo2018TasNet:}, and WaveGAN \cite{Donahue2018Adversarial} have shown that using a large temporal context window is important in waveform modeling. To efficiently take advantage of the long-range dependency of speech signals, dilated convolution was proposed in \cite{Yu2015Multi-Scale}. In \cite{Rethage2017Wavenet, Oord2016WaveNet:,Yu2015Multi-Scale}, the effectiveness of the dilated convolutional layers was shown to expand the receptive field exponentially (rather than linearly) with depth. Fig. 2 shows an example that demonstrates the concept of dilated fully convolutional filters. The input signal (I) is processed by a dilated convolutional block to generate the output signal (O). 

The input sequence has 18 points. When using a one-dimensional fully convolutional filter to process the input signal, the number of receptive fields is 18. On the other hand, when using a dilated fully convolutional block with filter sizes of 2, 3, and 3 and dilated rates of 1, 2, and 6, the receptive field is also 18. Compared to a single-layered FCN block, with the same size of receptive fields, the dilated fully convolutional block requires only half the number of parameters but four times the depth, suggesting that the dilated fully convolutional block can have a deeper architecture than the conventional fully convolutional filter when the total number of parameters is fixed.

\begin{figure}
  \centering
  \includegraphics[width=0.4\textwidth]{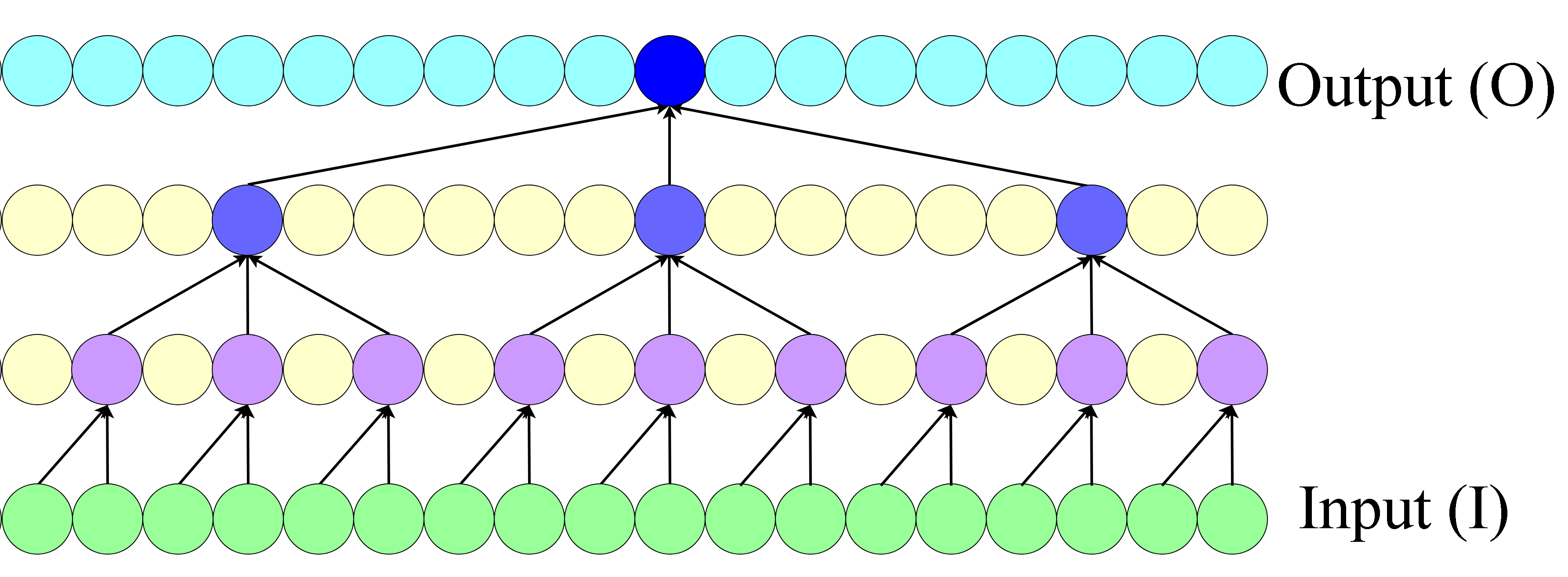}
  \caption{Input (I) and output (O) with two-layered dilated convolutional filters.}
\end{figure}

\section{THE PROPOSED MULTICHANNEL SE SYSTEM}
In this section, we first introduce the proposed SDFCN multichannel SE system. Then, we explain the extended system, rSDFCN. The design concept and architectures of SDFCN and rSDFCN are presented in detail.

\begin{figure}
  \centering
  \includegraphics[width=0.3\textwidth]{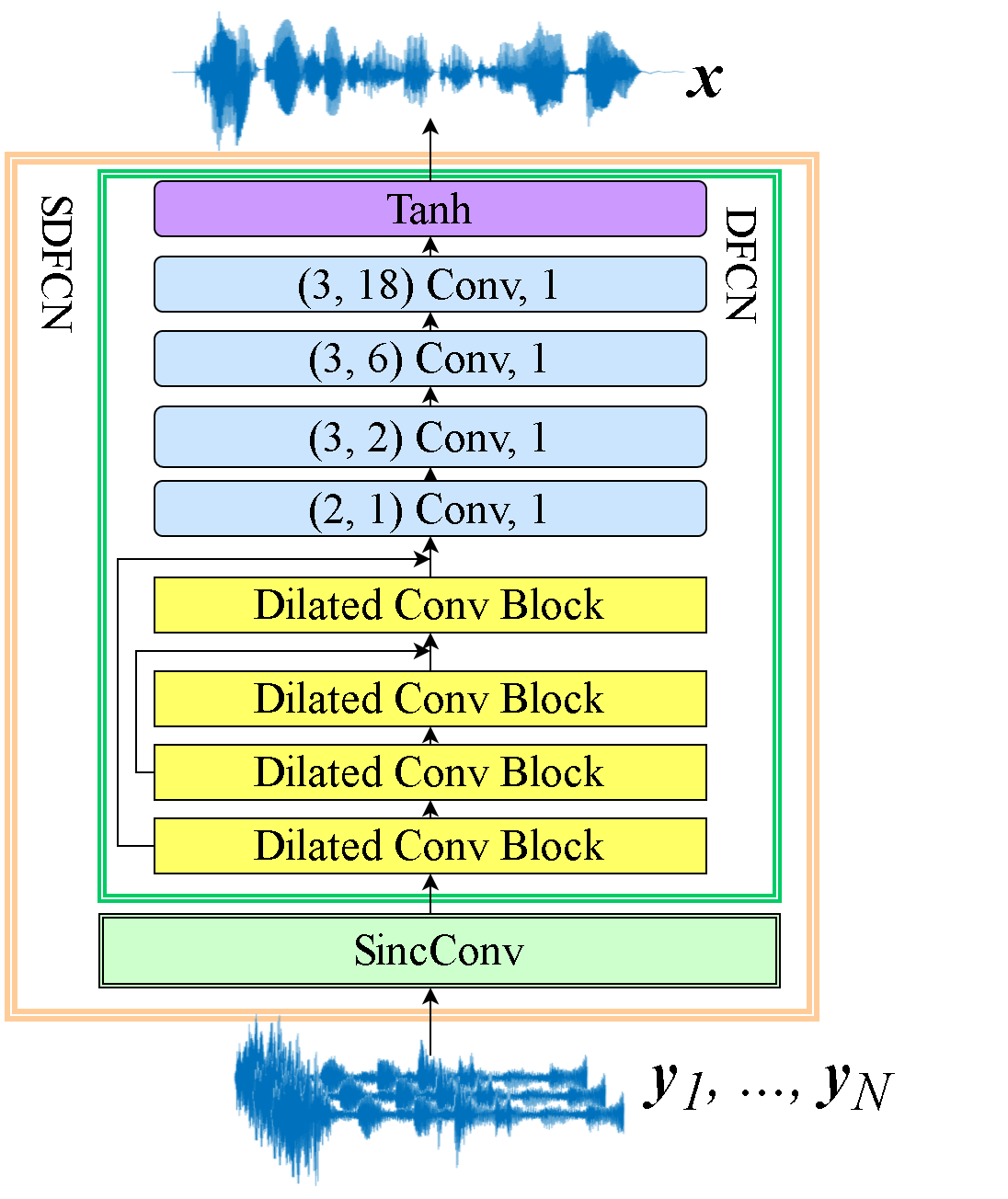}
  \caption{Architecture of the SDFCN multichannel SE system. Each of four blue rectangles denotes one dilated convolutional layer, and the parameters are denoted as follows: ($p1$, $p2$) Conv $p3$, where $p1$ is the kernel size, $p2$ is the dilated rate, \textcolor{ColorVariable2}{and $p3$ is the number of filters (channels).}}
\end{figure}

\begin{figure}
  \centering
  \includegraphics[width=0.3\textwidth]{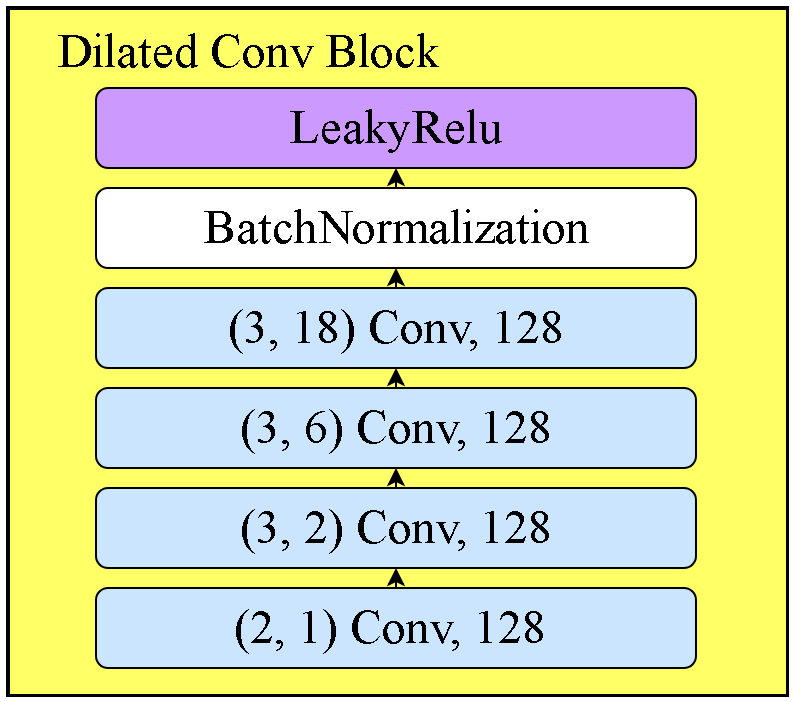}
  \caption{Architecture of the dilated convolutional block in the SDFCN model.}
  \label{fig:f4}
\end{figure}

\subsection{The SDFCN System}
Fig. 3 shows the architecture of the proposed SDFCN multichannel SE system, which consists of a SincConv layer and a dilated FCN (termed DFCN) module. The DFCN module consists of four layers of dilated convolutional blocks (Dilated Conv Block in Fig. 3), four dilated convolutional layers, and a tanh activation function layer. A skip-connection scheme is adopted to provide additional low-level information to the higher-level process. From our preliminary experimental results, we note that with such a skip-connection scheme, the SDFCN model can be trained more efficiently. Given the multichannel inputs: $\mathbf{Y}=[\mathbf{y}_1,\mathbf{y}_2,\dots,\mathbf{y}_N]$, where $N$ denotes the number of channels, we have 
\begin{equation}
	\mathbf{\hat{x}}= f_{DFCN} ( f_{SincConv} (\mathbf{Y})),
\end{equation}
where $f_{SincConv}(\cdot)$ and $f_{DFCN}(\cdot)$ denote the mapping functions of the SincCnov layer and the DFCN module, respectively. Fig. 4 shows the architecture of the dilated convolutional blocks (Dilated Conv Block in Fig. 3) in the SDFCN model. The block consists of four dilated convolutional layers (the four blue rectangles) followed by batch normalization and LeakyRelu. \textcolor{ColorVariable2}{The receptive field of the dilated convolutional block is 54 ($2\times3\times3\times3$), which is designed to approximate the kernel size of a Conv layers in FCN \cite{Fu2018End-to-End}.} 

\begin{figure}
  \centering
  \includegraphics[width=0.3\textwidth]{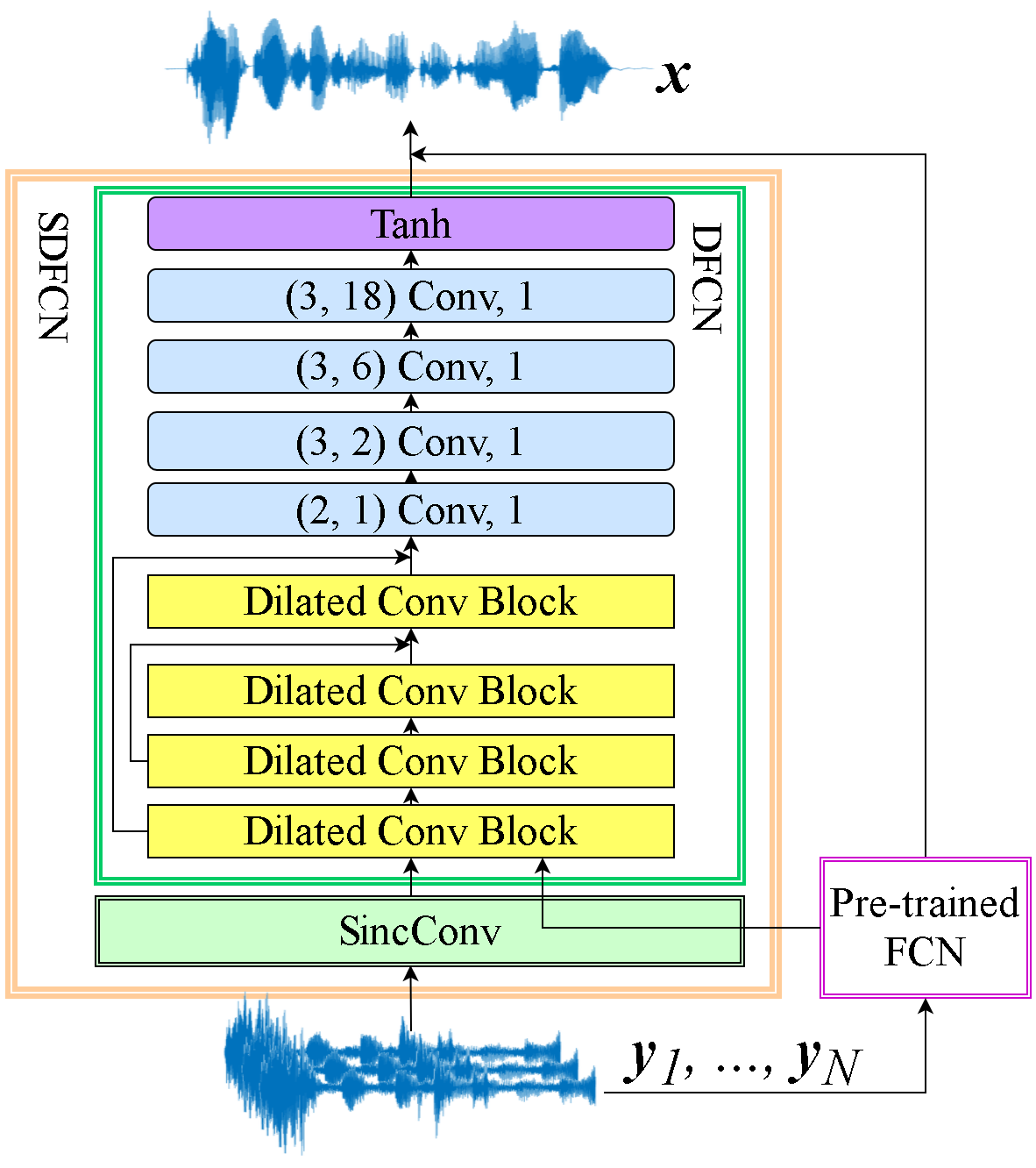}
  \caption{Architecture of the rSDFCN multichannel SE system, which consists of a primary SE module (pre-trained FCN) and a SDFCN system.}
\end{figure}

\subsection{The Residual SDFCN (rSDFCN) System}
Recently, residual structures have been popularly used in neural network models to attain better classification and regression efficacy. \textcolor{ColorVariable2}{In speech signal generation tasks, residual connections also provide promising performance because the residual connection provides a linear shortcut, and the non-linear part of the network only needs to deal with the residuals (differences) of the estimated and reference signals, which are usually easier to model. In this work, we also explore the combination of the residual structures with SDFCN. This combined model is termed the residual SDFCN (rSDFCN). The architecture of an rSDFCN multichannel SE system is shown in Fig. 5.} \par
As can be seen from the figure, an additional SE module (the pre-trained FCN in Fig. 5) is used. This SE module is treated as the primary SE module, and the output of the primary SE module is combined with the output of the SDFCN system to form the final enhanced output. The formulation of the rSDFCN can be represented as:

\textcolor{ColorVariable1}
{\begin{equation}
	\mathbf{\hat{x}}= f_{DFCN} ( f_{SincConv} (\mathbf{Y}),f_{Pr} (\mathbf{Y}))+f_{Pr} (\mathbf{Y}),
\end{equation}}

\noindent where $f_{Pr}(\cdot)$ is the mapping function of the primary SE module.
When implementing the rSDFCN system, we first pre-train the primary SE module and then train the SDFCN system. In this way, the SDFCN system learns the residual components (or differences) of the clean reference and the enhanced output of the primary SE module. More specifically, the SDFCN system is trained with the aim of minimizing the following loss function: 

\textcolor{ColorVariable1}{
\begin{equation}
	\|f_{DFCN} ( f_{SincConv} (\mathbf{Y}),f_{Pr}(\mathbf{Y}))-[\mathbf{x}-f_{Pr} (\mathbf{Y})]\|^2.
\end{equation}}

In this paper, we use a pre-trained FCN model as the primary SE module. Its architecture is shown in 
Fig. 6. The module consists of seven layers of convolution blocks, a convolutional layer, and a tanh activation function layer. Each convolution block consists of a convolutional layer (with length = 55 and channel = 64), batch normalization, and LeakyRelu. {\color{ColorVariable3} Please note that the architectures of the FCN, SDFCN, and rSDFCN presented above are designed based on the datasets used in this study. The parameters, including the numbers of layers and channel filters and the kernel size can be adjusted according to the target task.}   

\begin{figure}
  \centering
  \includegraphics[width=0.3\textwidth]{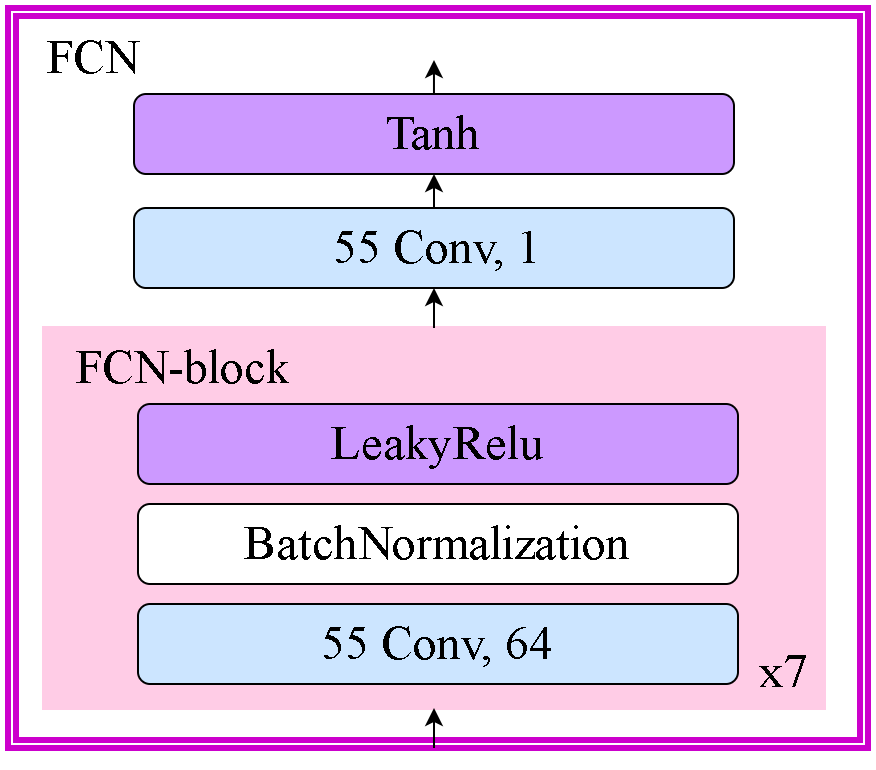}
  \caption{Architecture of the FCN model that is used as the primary SE module in the proposed rSDFCN system. We use $p_1$ Conv $p_2$ to represent a convolutional layer with $p_2$ filters and kernel size of $p_1$.}
  \label{fig:f6}
\end{figure}

\section{EXPERIMENTAL SETUP AND RESULTS}
In this section, we first introduce the experimental setup for the IEM-SE and DM-SE tasks{\footnote{Speech samples and codes can be found via: https://yu-tsao.github.io/MCSE/}}. Then, we present the results of the proposed SDFCN and rSDFCN systems for these two tasks. \textcolor{ColorVariable1}{Finally, we discuss the performance of the rSDFCN system on several subsets of the CHiME-3 dataset with different subset size. For IEM-SE and CHiME-3 task, we also discuss the effectiveness of dilated convolution and SincConv layer.} 

\subsection{Experimental Setup}
We evaluated the SE performance in terms of two standard objective metrics: STOI \cite{Taal2010short-time,Taal2011Algorithm} and PESQ \cite{RixPerceptual}. The STOI score ranges from 0 to 1, and the PESQ score ranges from 0.5 to 4.5. For STOI and PESQ, a higher score indicates that the enhanced speech signal has higher intelligibility and better quality, respectively, with reference to the speech signal recorded by the near-field high-quality microphone. In addition, we also conducted listening tests and evaluated the speech recognition performance of enhanced speech in terms of the Chinese character error rate (CER) using Google Speech Recognition \cite{Zhang2017Speech}. 
For comparison, we implemented a DDAE-based multichannel SE system \cite{Lu2014Ensemble,Lu2013Speech}. In previous studies, the single-channel DDAE approach has shown outstanding performance in noise reduction \cite{Lai2018Deep}, dereverberation \cite{Lee2018Speech}, and bone-conducted speech enhancement [58]. Here, we extended the original single-channel DDAE approach to form a multichannel DDAE system. Fig. 7 shows the architecture of the multichannel DDAE system, which consists of five dense layers. The input is multiple sequences of noisy spectral features (log-power spectrogram (LPS) in this study) from the multiple channels, and the output is a sequence of enhanced spectral features. The phase of one of the noisy speech utterances was used as the phase to reconstruct the enhanced waveform. All neural network models were trained using the Adam optimizer \cite{Kingma2014Adam:} with a learning rate of 0.001. The $\alpha$ value of LeakyReLU was set to 0.3.

\begin{figure}
  \centering
  \includegraphics[width=0.3\textwidth]{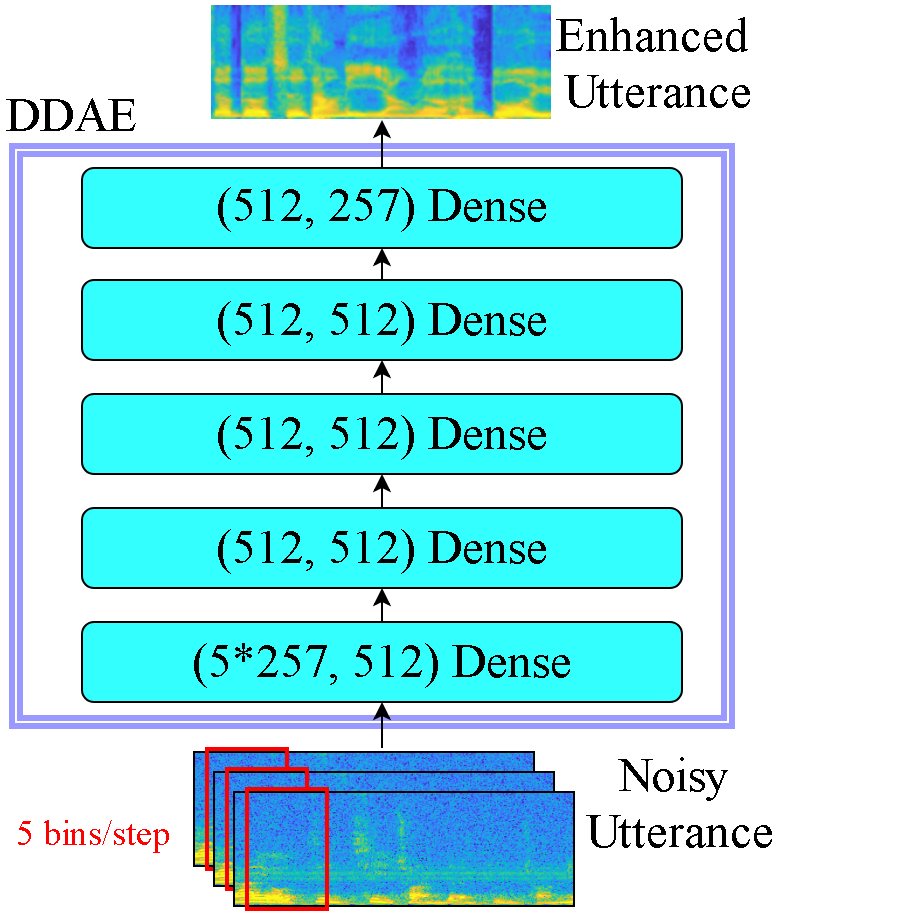}
  \caption{Architecture of the DDAE multichannel SE system.}
\end{figure}

\subsection{\textcolor{ColorVariable1}{The Inner-Ear Microphones-SE Task}}
When speech signals are recorded using inner-ear microphones, interference from the environment can be blocked, so that purer signals can be captured. However, owing to the different transmission pathways, the speech signals captured by the IEMs exhibit different characteristics from those recorded by normal air-conducted microphones (ACMs). Generally speaking, the high-frequency components of speech recorded by an IEM are suppressed, thereby notably degrading the speech quality and intelligibility. Moreover, owing to the loss of high-frequency components, IEM speech cannot provide a satisfactory ASR performance.
 
For the IEM-SE task, we intend to transform the speech signals captured by a pair of IEMs into ACM-like speech signals with improved quality and intelligibility. In the past, there have been some studies on IEM-to-ACM transformation. In \cite{Kondo2006On,Bouserhal2017In-ear}, bandwidth expansion and equalization techniques were used to map the IEM speech signals to the ACM ones. Because the mapping function between IEM and ACM is nonlinear and complex, traditional linear filters may not provide optimal performance. In the present study, we propose to perform multichannel SE in the waveform domain for IEM-to-ACM transformation. 

\begin{figure}
  \centering
  \includegraphics[width=0.45\textwidth]{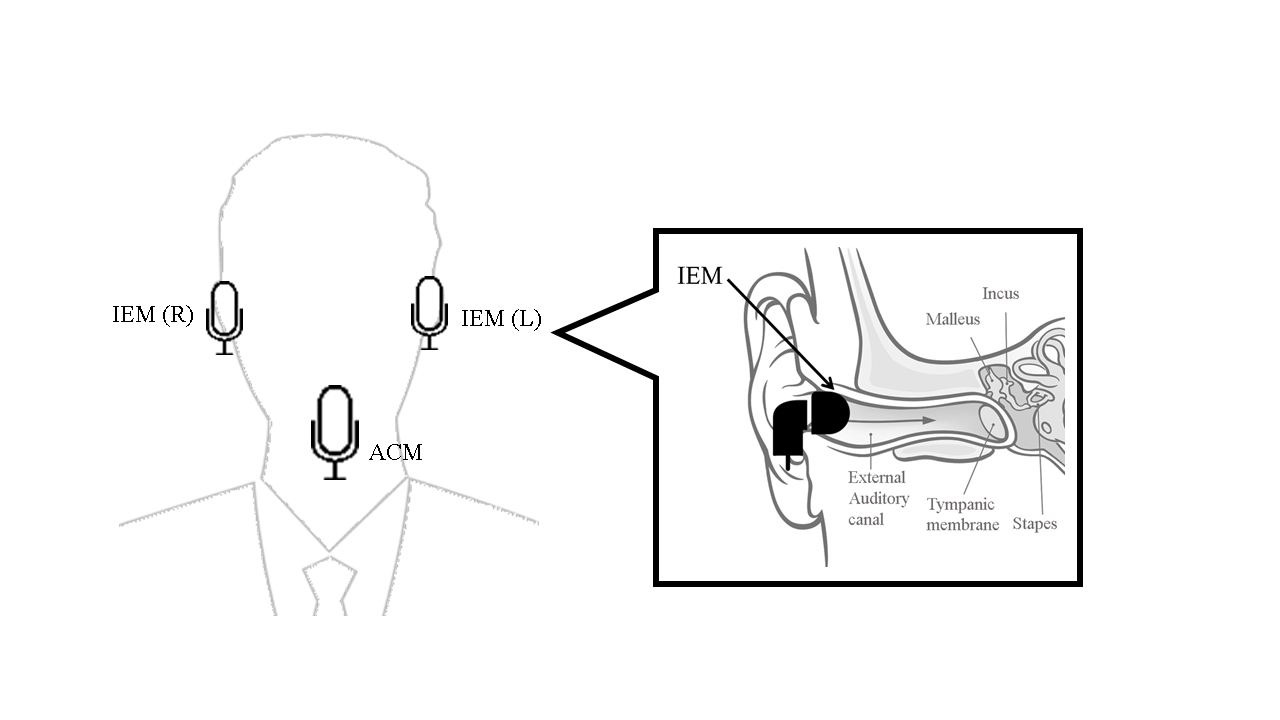}
  \caption{Recording setting of IEM-SE task. There is a near-mouth microphone and two IEMs in both ears.}
  \label{fig:f8}
\end{figure}

Our recording condition is shown in Fig. \ref{fig:f8}. A male speaker sat in a \textcolor{ColorVariable1}{ sound booth (3m$\times$5.2m, 2m in height)} and wore a pair of IEMs and a near-mouth ACM. The three microphones simultaneously recorded speech signals spoken by the male speaker. The recording scripts were the Taiwan Mandarin Chinese version of Hearing in Noise Test (TMHINT) sentences \cite{Wong2007Development}. There were 250 utterances for training and another 50 utterances for testing. {\textcolor{ColorVariable3}All utterances were recorded at a 16,000 Hz sampling rate then truncated to speech segments, each containing 36,500 sample points (around 2.28 seconds).}

Before discussing the results of the proposed SDFCN and rSDFCN systems, we first verified the effectiveness of dilated convolution and SincConv layer. There were totally four models trained in this preliminary experiment. We compared \textit{FCN-55} with \textit{DCN-54} to show the effect of dilated convolution. The benefits of SincConv layer were shown by comparing \textit{FCN-251} with \textit{SincFCN-251}.  \textit{FCN-55} is similar to regular FCN shown in Fig. \ref{fig:f6}, but it has only four layers in total. \textit{DCN-54} was designed by replacing the last three Conv layers in \textit{FCN-55} with  dilated convolutional block shown in Fig. \ref{fig:f4}, where the kernel size is 55. The reason for using four-layer models is that models with less than four layers could not enhance utterances well in our preliminary experiments. As mentioned in the previous section, the receptive field of the dilated convolutional blocks was set to be 54 to approximate the kernel size used in FCN. \textit{FCN-251} was designed by changing the kernel size of the first Conv layer in \textit{FCN-55} from 55 to 251, and \textit{SincFCN-251} was designed by replacing the first Conv layer in \textit{FCN-251} with a SincConv layer. The reason that we changed the kernel size of the first layer was to make it have the same size as the original work \cite{Ravanelli2018Speaker}. \textcolor{ColorVariable2}{For a fair comparison, the numbers of filters of all models trained and tested in the experiment are 30. }

\textcolor{ColorVariable1}{
Table \ref{table:1} lists the average STOI and PESQ scores of the original speech signals captured by the left and right IEMs (denoted as IEM (L) and IEM (R), respectively) and the enhanced speech signal by the four models mentioned above. The corresponding ACM speech was used as the reference to compute the scores. By comparing the results of the middle columns in Table \ref{table:1}, we observe that the STOI and PESQ scores can be further improved by the dilated convolutional layer.  The results in the last column in Table \ref{table:1} show that the SincConv layer performs much better than the original convolutional layer.}
\begin{table*}
\color{ColorVariable1}
\centering
\caption{\textcolor{ColorVariable1}{Average STOI and PESQ Scores of \textit{FCN-55}, \textit{DCN-54}, \textit{FCN-251}, \textit{SincFCN-251}, and Single-channel/Multichannel SDFCN Models for IEM-SE Task.}}
\begin{tabular}{|c | c c | c c | c c |} 
 \hline
 No. of ch. & 1 & 1 & 2 & 2 & 2 &2\\
 \hline
 Model & IEM(L) & IEM(R) & \textit{FCN-55} &\textit{DCN-54} & \textit{FCN-251} & \textit{SincFCN-251}\\
 \hline
 STOI & 0.694 &	0.694 & 0.801 & \textbf{0.817} & 0.727  & \textbf{0.843}\\
 \hline
 PESQ &	1.146 &	1.101 &	1.317 &	\textbf{1.360} & 1.171  & \textbf{1.476}\\
 \hline
 \end{tabular}
\label{table:1}
\end{table*}

\begin{figure}
  \centering
  \includegraphics[width=0.45\textwidth]{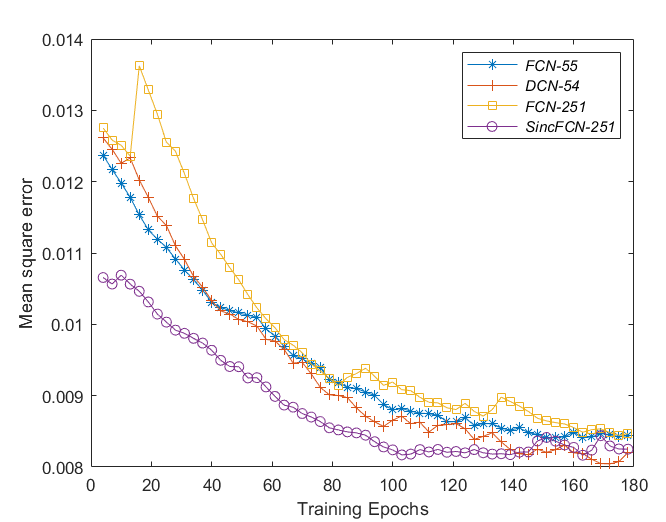}
  \caption{\textcolor{ColorVariable1}{MSE on the test set of \textit{FCN-251}, \textit{DCN-54}, \textit{FCN-251}, and \textit{SincFCN-251} over training epochs.}}
  \label{fig:curve}
\end{figure}

\begin{table}\sffamily
\begin{tabularx}{0.47\textwidth}{ m{1.1em}  m{7cm}} 
(a) & \includegraphics[width=\linewidth,height=3.6cm]{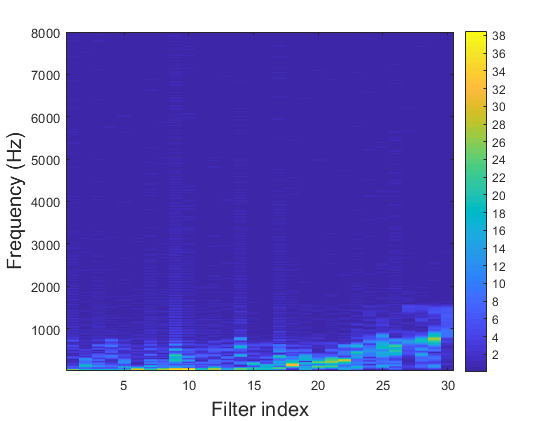} \\
(b) & \includegraphics[width=\linewidth,height=3.6cm]{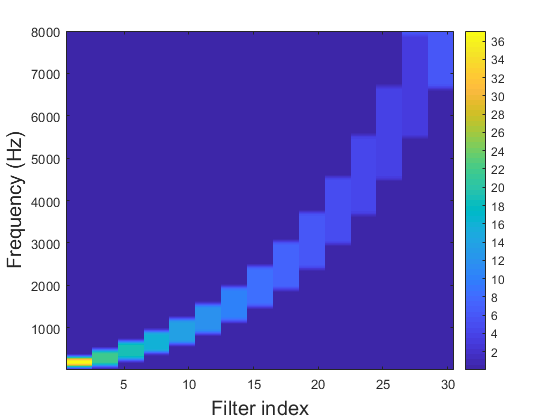} 
\end{tabularx}
\captionof{figure}{
\textcolor{ColorVariable1}{
Frequency responses of the learned filters of (a) the first layer of \textit{FCN-251} and (b) the SincConv layer of \textit{SincFCN-251}.
}
}
\label{table:filters}
\end{table}

\begin{table}\sffamily
\centering
\begin{tabularx}{0.47\textwidth}{ m{4cm}  m{4cm}} 
\includegraphics[width=\linewidth,height=4.5cm]{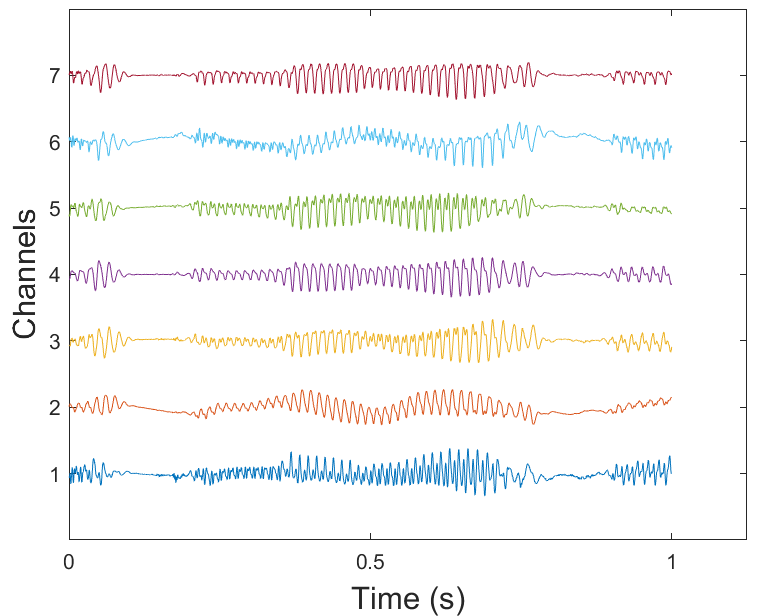}  & \includegraphics[width=\linewidth,height=4.5cm]{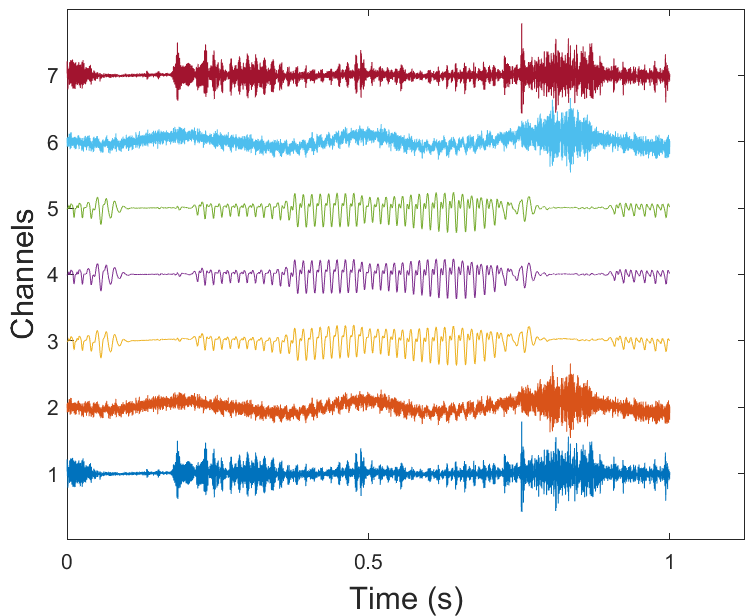} \\
\centering (a) &  \centering (b) 
\end{tabularx}
\captionof{figure}{
\textcolor{ColorVariable2}{
Extracted features of an utterance audio from convolution by the filters of the first layers of (a) \textit{FCN-251} and (b) \textit{SincFCN-251}.
}
}
\label{table:filters_feature}
\end{table}

Fig. \ref{fig:curve} shows the learning curves of the four models in terms of the MSE scores. When computing the MSE scores, we have pre-processed each utterance by normalizing the waveform samples by the peak amplitude. From Table \ref{table:1} and Fig. \ref{fig:curve}, we can see that although their losses (MSE) converge to a similar value, the training speed of \textit{SincFCN-251} is much faster, and the corresponding STOI and PESQ scores are also higher than others. It is also noted that, \textit{DCN-54} and \textit{SincFCN-251} outperform \textit{FCN-55} and \textit{FCN-251} in terms of STOI and PESQ, respectively, which confirms the effectiveness of the dilated convolution block and SincConv layer. Also, from Table \ref{table:1}, we can observe that \textit{FCN-55} outperforms \textit{FCN-251}, which implies that the performance of FCN can not be improved by just increasing the kernel size of the convolutional layers. In Table \ref{table:dilatedMSE}, we further list the average MSE scores of \textit{FCN-55}, \textit{DCN-54}, \textit{FCN-251}, and \textit{SincFCN-251} under the 180 training epoch condition in Fig. \ref{fig:curve}. The results in the table show that \textit{DCN-54} and \textit{SincFCN-251}, respectively, yield lower MSE scores as compared to \textit{FCN-55} and \textit{FCN-251}, again confirming the benefits of the dilated convolution and SincConv. 
\par

\begin{table}
\color{ColorVariable2}
\centering
\caption{\textcolor{ColorVariable2}{Average MSE Scores of \textit{FCN-55}, \textit{DCN-54}, \textit{FCN-251}, and \textit{SincFCN-251} for IEM-SE Task.}}
\begin{tabular}{|c| c c | c c |} 
 \hline
 No. of ch. & 2 & 2 & 2 &2\\
 \hline
 Model & \textit{FCN-55} &\textit{DCN-54} & \textit{FCN-251} & \textit{SincFCN-251}\\
 \hline
 MSE  & 0.00844 & \textbf{0.00821} & 0.00847  & \textbf{0.00825}\\
 \hline
 \end{tabular}
\label{table:dilatedMSE}
\end{table}

\textcolor{ColorVariable2}{To visually compare FCN and SincConv, we plot the learned filters of \textit{FCN-251} and \textit{SincFCN-251} in Fig. \ref{table:filters} {\color{ColorVariable3}(for a clearer presentation, we only used 30 filters for both FCN and SincConv to plot Fig. \ref{table:filters})}. In the meanwhile, we plot the extracted features of an utterance from FCN and SincConv layers in Fig. \ref{table:filters_feature} (for a clearer presentation, we only used seven filters for both {\color{ColorVariable3} \textit{FCN-251} and \textit{SincFCN-251}} to obtain the features in Fig. \ref{table:filters_feature}). From Figs. \ref{table:filters} and \ref{table:filters_feature}, and Table \ref{table:dilatedMSE}, we can note that our experiment results are quite consistent with those in the previous works \cite{Ravanelli2018Speaker, mittermaier2019small,gong2019dilated}. From Fig. \ref{table:filters}, we can see that the SincConv layer learns a filter bank containing more filters with high cut-frequencies compared to the traditional convolutional layer. The filters learned by FCN, as shown in Fig. \ref{table:filters}(a), do not cover all the frequency ranges. From Fig. \ref{table:filters_feature}, we can observe that the first-layered features of \textit{SincFCN-251} contain more high frequency components than \textit{FCN-251}. In addition, the results in Table \ref{table:dilatedMSE} are consist with those in \cite{gong2019dilated}: With dilated convolution, the network can more accurately model ground-truth waveforms in terms of MSE. } \par

\textcolor{ColorVariable1}{Furthermore, to investigate the effectiveness of using multiple (dual) channels, we also compared the SDFCN model trained with dual-channel input and that trained with single-channel input. The results are denoted as SDFCN (using dual-channel inputs), SDFCN(L) (using the left channel only) and SDFCN(R) in the left part of Table \ref{table:2}}. From the table, we first note that SDFCN(L) and SDFCN(R) achieve improved STOI and PESQ scores over IEM(L) and IEM(R), respectively. The results confirm the effectiveness of the proposed SDFCN system for single microphone SE. Next, we note that SDFCN outperforms both SDFCN (L)and SDFCN(R), confirming the advantage of the multichannel (dual-channel) mode over its single-channel counterparts. 

\begin{table*}
\centering
\caption{\textcolor{ColorVariable1}{Average STOI and PESQ Scores of Different Single-channel/Multichannel SE Models for the IEM-SE Task.}}
\begin{tabular}{|c | c c c | c c c c|} 
 \hline
 No. of ch.& 1 & 1 & 2 & 2 & 2 & 2 & 2 \\
 \hline
 Model &  SDFCN(L) & SDFCN(R) & SDFCN & DFCN & FCN & DDAE & rSDFCN\\
 \hline
 STOI & 0.861 & 0.824 &	\textbf{0.880} & 0.867 & 0.834 & 0.773 & \textbf{0.894}\\
 \hline
 PESQ & 1.631 & 1.597 &	\textbf{1.643} & 1.562 & 1.446 & 1.939 & \textbf{1.986}\\
 \hline
 \end{tabular}
\label{table:2}
\end{table*}

\textcolor{ColorVariable1}{Next, we report the results of rSDFCN in the right part of Table \ref{table:2}}. To confirm the effectiveness of SincConv, we replaced the SincConv layer in SDFCN with a normal convolutional layer, denoted as DFCN. FCN denotes the results of the pre-trained FCN module used in rSDFCN. Comparing the results of SDFCN and DFCN in Table \ref{table:2}, we confirm the effectiveness of SincConv for the SE task. Comparing the results of SDFCN, FCN and rSDFCN in Table \ref{table:2}, we confirm the effectiveness of the residual architecture for the SE task. Next, we note that both SDFCN and rSDFCN outperform the baseline DDAE system while rSDFCN outperforms SDFCN. 
	
In addition to comparing the objective scores, we also conducted qualitative analysis. Fig. \ref{fig:f9}(a), (b), (c), (d), and (e) show the waveforms and spectrograms of the near-field ACM, IEM(L), and IEM(R) speech signals and the enhanced speech signals obtained by rSDFCN and DDAE, respectively. By comparing Fig. \ref{fig:f9}(a), (b), and (c), we can easily note that the IEM speech signals suffer from notable distortion, with high-frequency components being suppressed. Next, by comparing Fig. \ref{fig:f9} (a) and (d), we note that the proposed rSDFCN multichannel SE approach can generate an enhanced speech signal similar to the ACM recorded speech signal. We can also observe that the DDAE-enhanced speech signal has a clearer structure in the high-frequency components while exhibiting some distortion in the low-frequency components. 

\begin{table}\sffamily
\begin{tabularx}{0.49\textwidth}{ m{1.1em}  m{3.5cm} m{3.5cm}} 
\toprule
 &  \:\:\:\:\:\:\:\:\:\:\:\:\:\:\:\:\:Waveform 
 &  \:\:\:\:\:\:\:\:\:\:\:\:\:\:\:Spectrogram \\ 
\midrule
(a) & \includegraphics[width=0.2\textwidth]{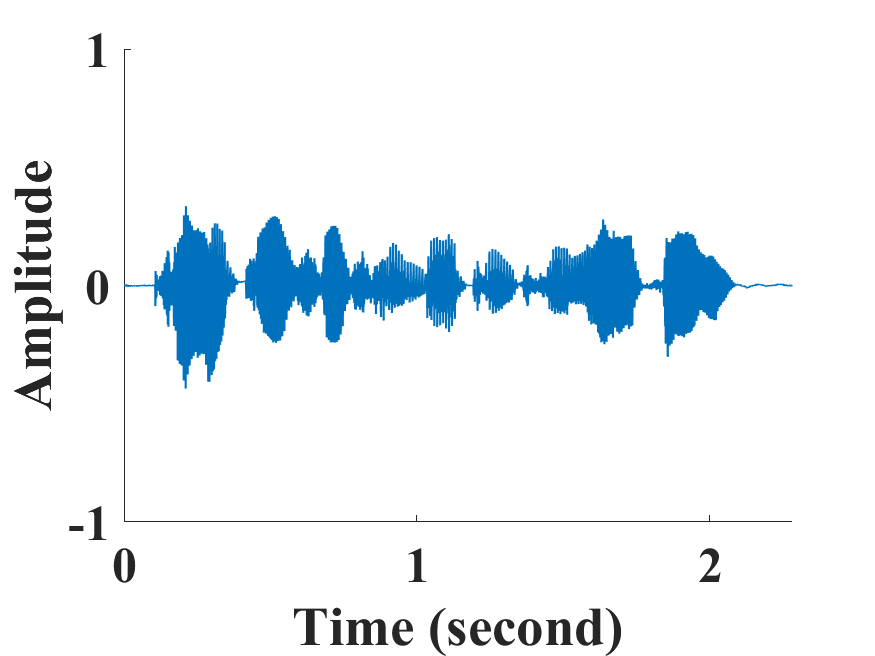} & \includegraphics[width=0.2\textwidth]{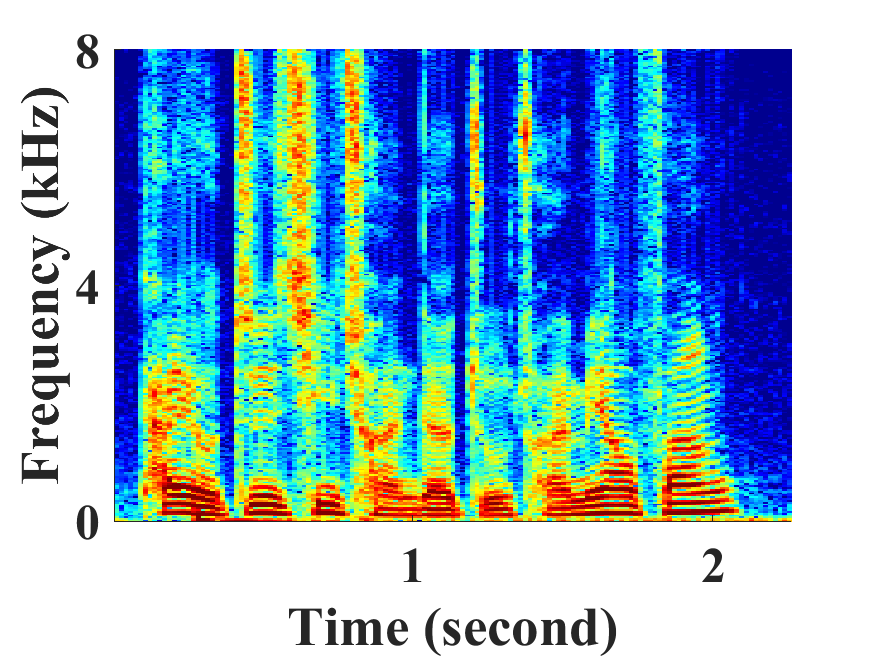}  \\ 
(b) & \includegraphics[width=0.2\textwidth]{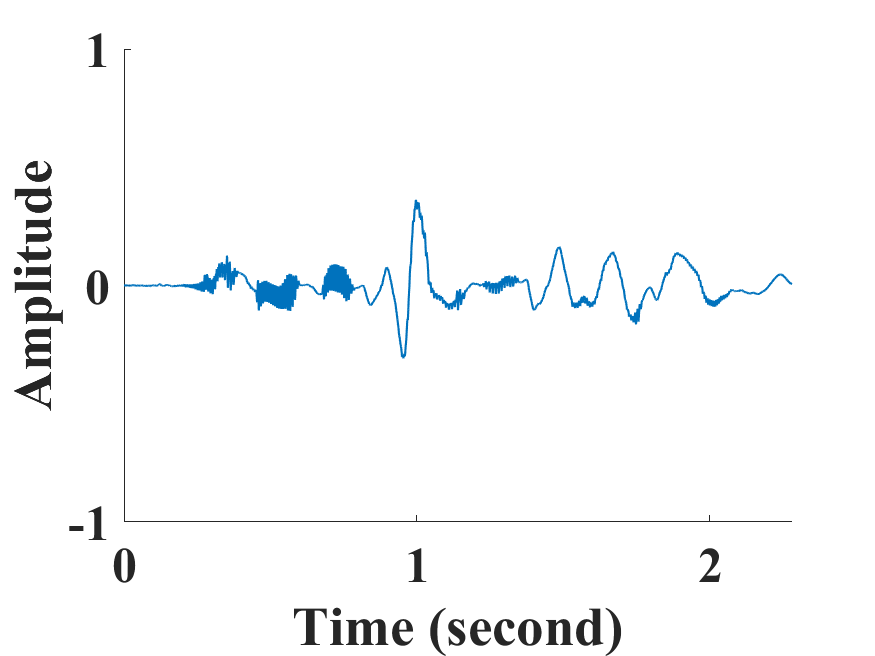} & \includegraphics[width=0.2\textwidth]{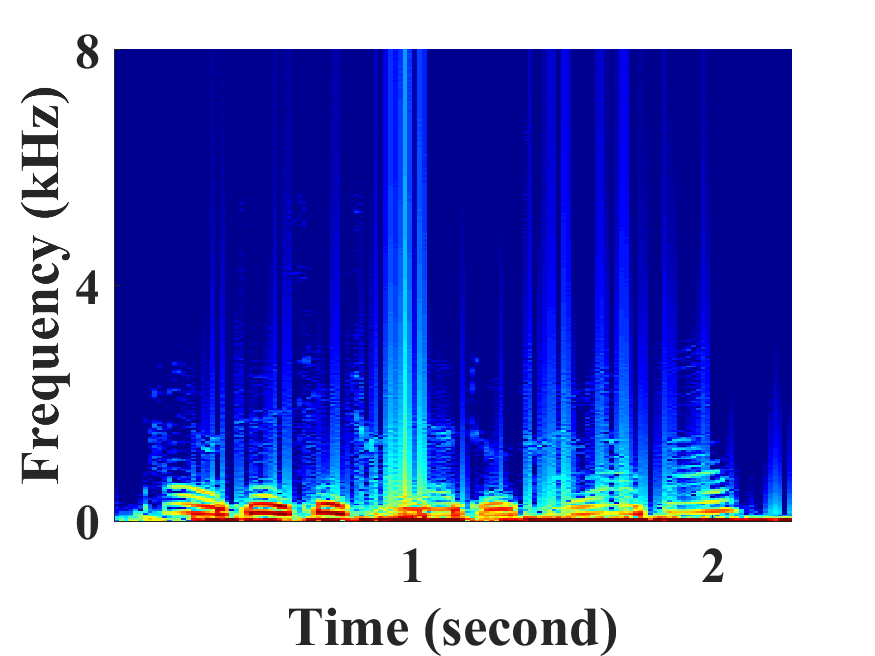}  \\ 
(c) & \includegraphics[width=0.2\textwidth]{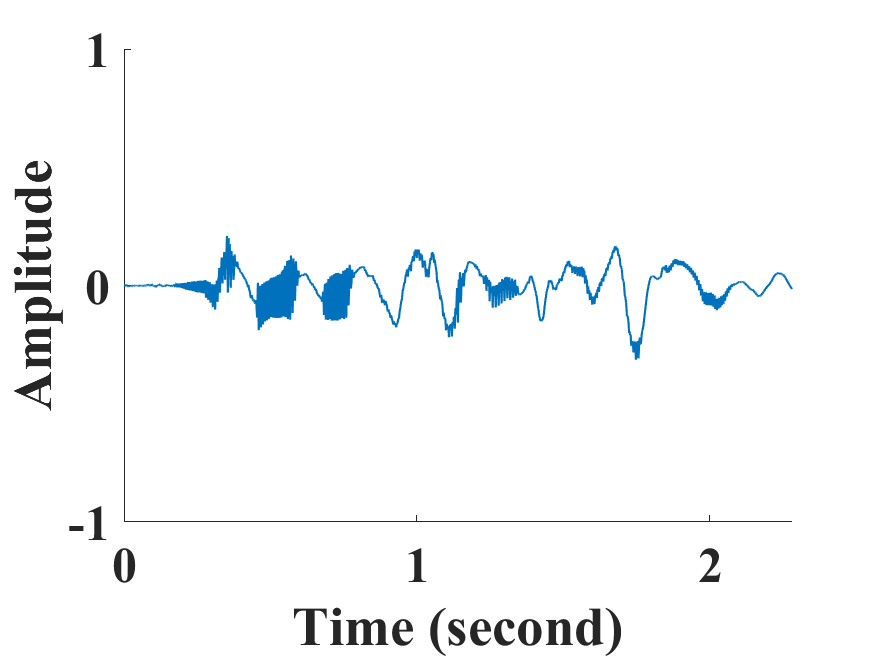} & \includegraphics[width=0.2\textwidth]{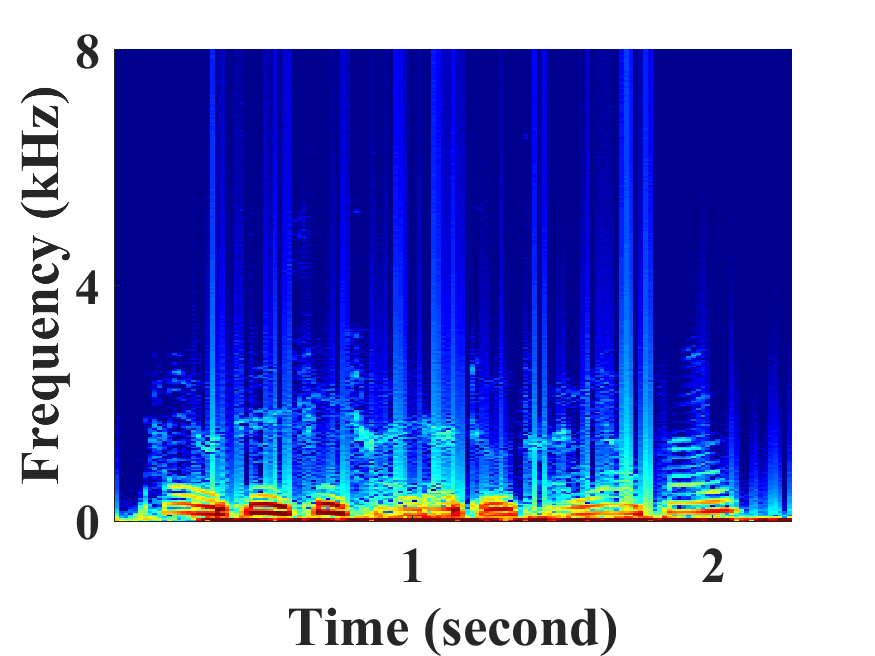}  \\ 
(d) & \includegraphics[width=0.2\textwidth]{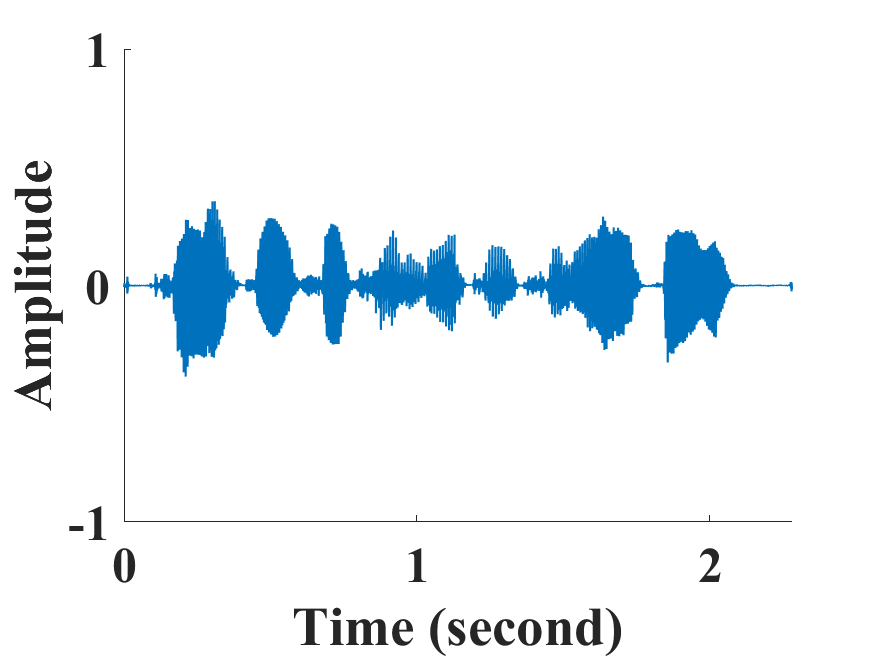} & \includegraphics[width=0.2\textwidth]{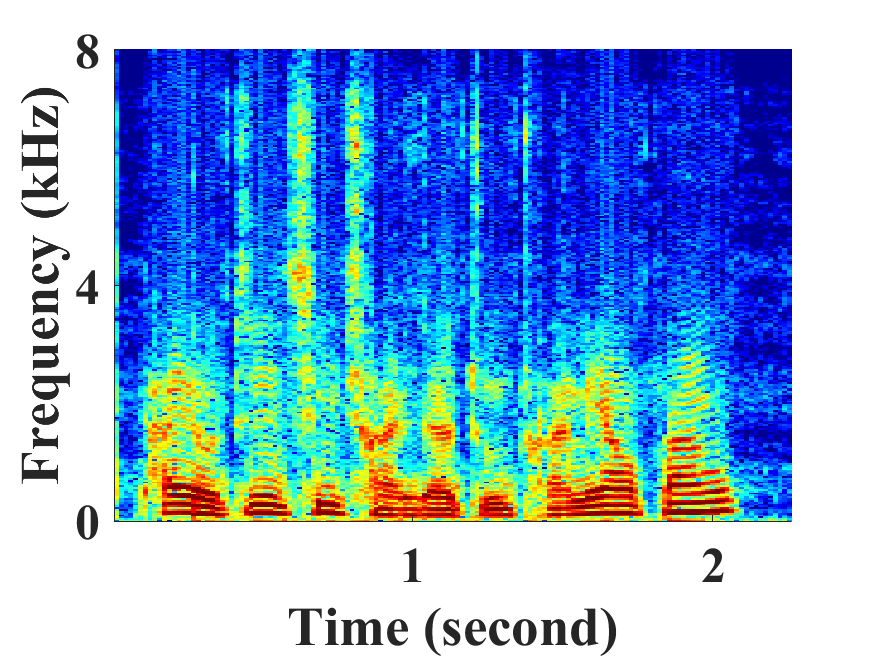}  \\ 
(e) & \includegraphics[width=0.2\textwidth]{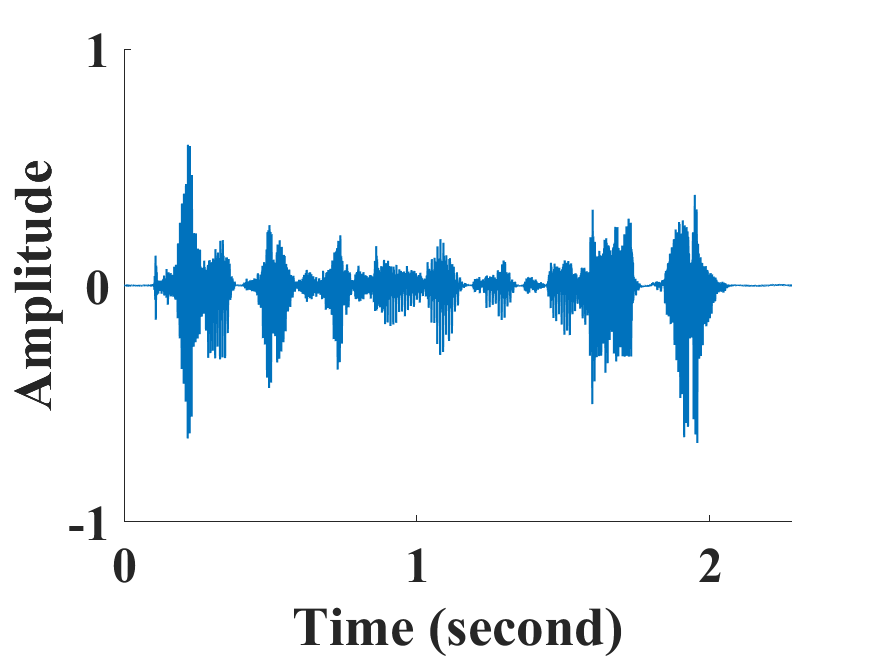} & \includegraphics[width=0.2\textwidth]{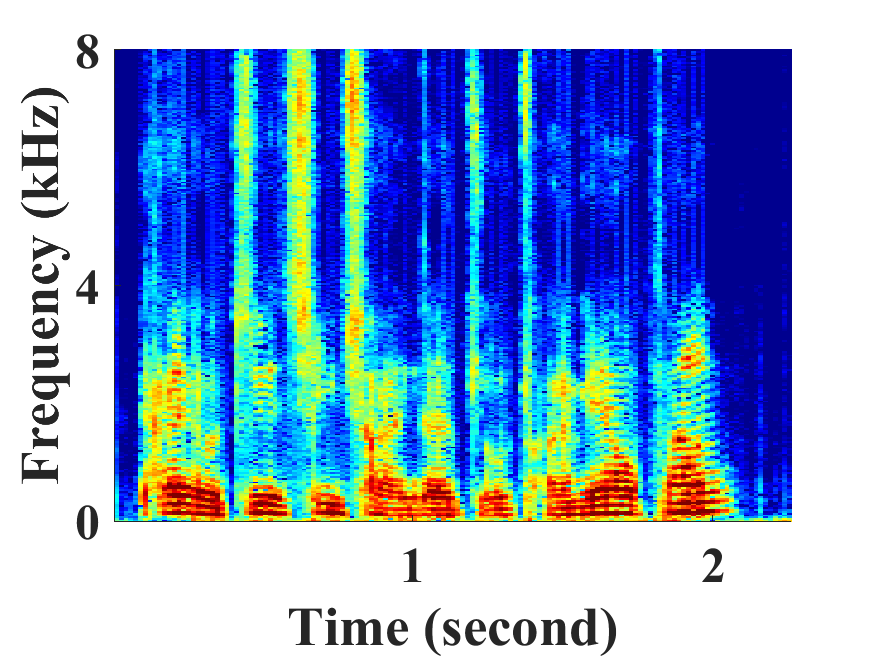}  \\ 
\bottomrule 
\end{tabularx}
\captionof{figure}{Waveforms and spectrograms of an example utterance in the IEM-SE task: (a) recorded speech by near-mouth microphone; (b) and (c) recorded speech by right and left IEMs, respectively. (d) and (e) enhanced speech by rSDFCN and DDAE, respectively.}
\label{fig:f9}
\end{table}

To subjectively evaluate the perceptual quality of the enhanced speech, we conducted AB reference tests to compare the proposed rSDFCN with the original IEM speech (here IEM(L) was used since it gave slightly higher PESQ scores in Table I). For comparison, the DDAE enhanced speech was also involved in the preference test. Accordingly, three pairs of listening tests were conducted, namely rSDFCN versus IEM, DDAE versus IEM, and rSDFCN versus DDAE. Each pair of speech samples were presented in a random order. For each listening test, speech samples were randomly selected from the test set. 15 listeners participated in the listening test. Listeners were instructed to select the speech sample with better quality. The stimuli were played to the listeners in a quiet environment through a set of Sennheiser HD headphones at a comfortable listening level. The results of the AB reference tests are presented in Fig. 13. From the figure, it is clear that rSDFCN and DDAE outperform IEM with notable margins, confirming the effectiveness of these two SE approaches. Next, we note that rSDFCN yields a higher preference score compared to DDAE, showing that rSDFCN can more effectively enhance the IEM speech.

\begin{figure}
  \centering
  \includegraphics[width=0.48\textwidth]{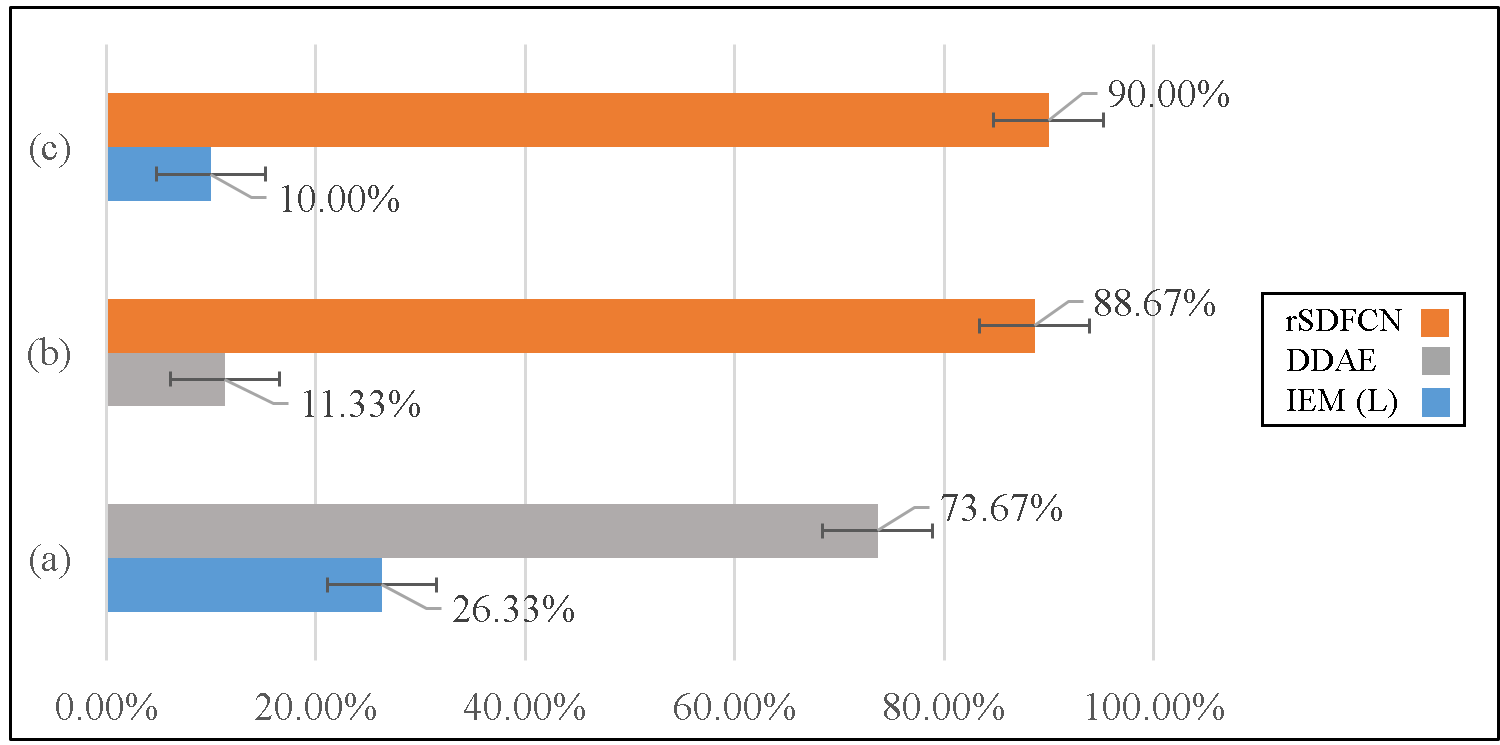}
  \caption{Results of the AB preference test (with 95\% confidence intervals) on speech quality compared between the proposed rSDFCN with IEM(L) and DDAE for the IEM-SE task.}
  \label{fig:s1}
\end{figure}

\begin{figure}
  \centering
  \includegraphics[width=0.3\textwidth]{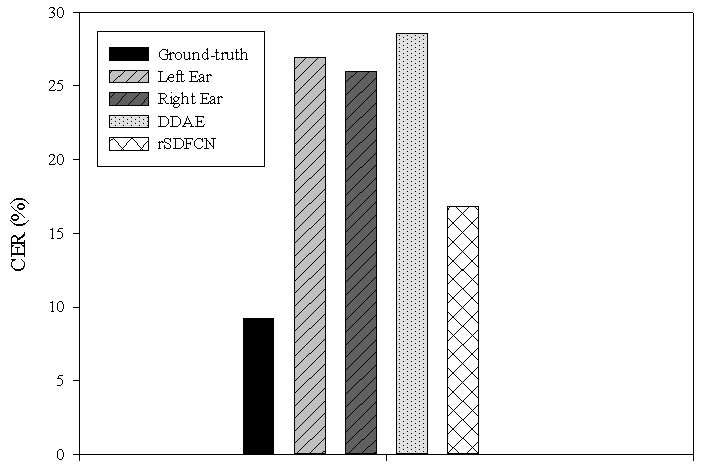}
  \caption{ASR results achieved by different SE models for the IEM-SE task.}
  \label{fig:f10}
\end{figure}

Finally, we tested the ASR performance in terms of the character error rate (CER). The results of the speech recorded by ACM, IEM(L), and IEM(R) and the enhanced speech by the rSDFCN and DDAE are shown in Fig. \ref{fig:f10}. The CER of the ACM-recorded speech is 9.2\%, which can be regarded as the upper-bound. The CERs of the speech recorded by IEM(L) and IEM(R) and the enhanced speech by rSDFCN and DDAE are 26.9\%, 26.0\%, 16.8\%, and 28.6\%, respectively. From the results, we note that rSDFCN can improve the ASR performance over IEM(L) and IEM(R). Compared with IEM(L), CER decreased by 35.38\% (from 26.0\% to 16.8\%). Comparing the results in Figs. \ref{fig:s1} and \ref{fig:f10} and Table III, we note that rSDFCN outperforms DDAE in terms of PESQ, STOI, subjective preference test scores, and ASR results, confirming the effectiveness of the proposed rSDFCN over the conventional DDAE approach for the IEM-SE task.

\subsection{\textcolor{ColorVariable1}{The  Distributed Microphone-SE Task}}

\begin{figure}
  \centering
  \includegraphics[width=0.3\textwidth]{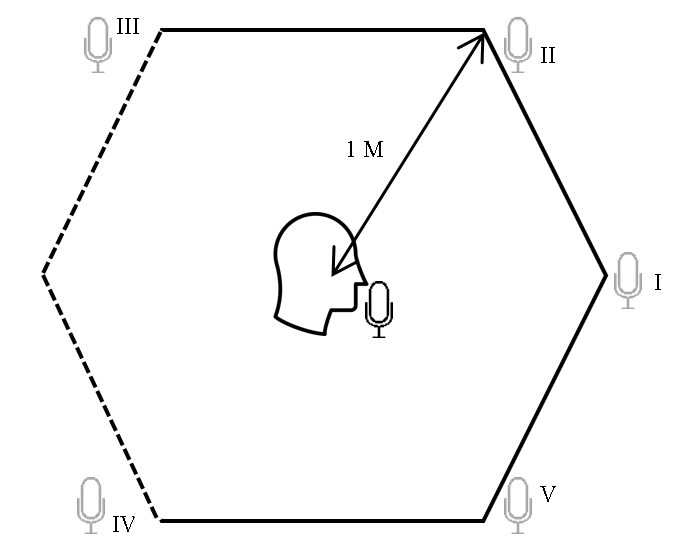}
  \caption{Recording setting for the DM-SE task. There is a near-field high-quality microphone and five far-field low-quality microphones. Distances of near-field microphone to far-field microphones are all 1 meter.}
  \label{fig:f11}
\end{figure}

For the DM-SE task, we also used the scripts of the TMHINT sentences to prepare the speech dataset. The layout of the recording is shown in Fig. \ref{fig:f11}. A high-quality near-field micro-phone (Shure PGA181 \cite{PGA181}) was placed right in front of the speaker and five low-quality microphones (all of the same brand and model: Sanlux HMT-11 \cite{SANLUX}) were located at the five vertices of the regular hexagon, 1 meter away from the speaker. \textcolor{ColorVariable1}{The room size is 15.5m$\times$11.2m and 3.27m in height.}  We labeled the low–quality microphones in counterclockwise order from I to V starting from the microphone in front of the speaker. 
Herein, the goal was to generate an enhanced (high-quality) speech signal using the speech signals recorded by the distant and low-quality microphones. To validate the effectiveness of using multiple channels for SE, we designed seven scenarios: five single-channel SE scenarios where the input consisted of the speech signal recorded by one of the five microphones [(I), (II), (III), (IV), or (V)] and the output was the enhanced speech signal, and two multichannel SE scenarios, where the input consisted of the speech signals recorded by three microphones (I, II, and V) and five microphones (I, II, III, IV, and V) and the output was the enhanced speech signal. For this set of experiments, we used 250 utterances for training and another 50 utterances for testing. \textcolor{ColorVariable3}{All utterances were recorded at 16,000 Hz and then truncated to speech segments, each containing 36,500 sample points (around 2.28 seconds).}
   
It is worth noting that although both IEM- and DM-SE tasks are multichannel SE scenarios, there are clear differences between them. For the IEM-SE task, the high-frequency components of the IEM speech signals are suppressed. In other words, the IEM speech resembles the low-pass-filtered ACM speech. Meanwhile, for the DM-SE task, the speech signals recorded by microphones I, II, III, IV, and V were degraded versions of the speech recorded by the near-field microphone owing to low-quality recording hardware, long-range fading, and room reverberation. As with the IEM-SE task, we tested the performance of rSDFCN and DDAE. 
	
\newcolumntype{P}[1]{>{\centering\arraybackslash}p{#1}}
\begin{table*}
\color{ColorVariable2}
\centering
\caption{\textcolor{ColorVariable2}{Average STOI and PESQ Scores of rSDFCN and DDAE for the DM-SE Task.}}
\begin{tabular}{|P{3cm}|| P{1.8cm}| P{1.8cm} |P{1.8cm} || P{1.8cm}| P{1.8cm} |P{1.8cm} |}
\hline
AVG. STOI/PESQ      & \multicolumn{3}{c|}{STOI}           & \multicolumn{3}{c|}{PESQ}               \\ \hline
Input Microphone(s) & Unenhanced & DDAE  & rSDFCN         & Unenhanced     & DDAE  & rSDFCN         \\ \hline
I                   & 0.872      & 0.823 & \textbf{0.932} & 1.602          & 1.618 & \textbf{1.648} \\ \hline
II                  & 0.896      & 0.814 & \textbf{0.930} & \textbf{1.736} & 1.606 & 1.656          \\ \hline
III                 & 0.888      & 0.813 & \textbf{0.931} & 0.526          & 1.623 & \textbf{1.644} \\ \hline
IV                  & 0.881      & 0.813 & \textbf{0.931} & 1.495          & 1.581 & \textbf{1.642} \\ \hline
V                   & 0.893      & 0.816 & \textbf{0.931} & \textbf{1.727} & 1.581 & 1.646          \\ \hline
I, II, V            &            & 0.823 & \textbf{0.950} &                & 1.655 & \textbf{1.780} \\ \hline
I, II, III, IV, V   &            & 0.829 & \textbf{0.954} &                & 1.635 & \textbf{1.826} \\ \hline
\end{tabular}
\label{table:3and4}
\end{table*}

Table \ref{table:3and4} show the average STOI and PESQ scores of rSDFCN and DDAE under seven conditions. The scores of the speech recorded by the far-field microphone (using the corresponding speech recorded by the near-field microphone as a reference) are also listed for comparison. From the tables, we can easily see that \textcolor{ColorVariable1}{rSDFCN} can improve the STOI and PESQ scores \textcolor{ColorVariable1}{when multichannel inputs are used.} When only one input is available (the task becomes a single-channel SE task), rSDFCN outperforms DDAE consistently across all of the five cases (single far-field microphone I, II, III, IV, and V). Meanwhile, for the multichannel task
({I, II, V} and {I, II, III, IV, V}), rSDFCN also outperforms DDAE. In addition, it is clear that the results of multichannel SE are superior to those of single-channel SE, implying that multichannel signals can provide useful information to more effectively enhance speech signals. \par
For qualitative analysis, the waveforms and spectrograms of a speech utterance recorded by the near-field microphone and the far-field microphone (channel II), along with the enhanced speech from rSDFCN and DDAE are shown in Fig. \ref{fig:f12}. For multichannel SE, we display the waveforms and spectrograms of the enhanced speech using five channels ({I, II, III, IV, and V}). \textcolor{ColorVariable2}{From Fig. \ref{fig:f12}(d) and (f), we can observe that DDAE provided a relatively clear structure of restored spectrogram, and rSDFCN outperformed DDAE when comparing the waveform plots in contrast. This result is reasonable because DDAE aims to minimize the MSE of spectral magnitude, while rSDFCN aims to minimize the MSE in the waveform domain. Because DDAE only enhances the magnitude spectrogram but not the phase information, it needs to borrow the phase information from the noisy speech when generating the speech waveforms. This may explain why DDAE performed worse than rSDFCN in terms of STOI and PESQ, as shown in Table \ref{table:3and4}, even though the spectrograms generated by DDAE were more similar to the ground-truth. This result is also consistent with those reported in previous works \cite{paliwal2011importance,le2011phase,gerkmann2015phase}.}\par

\newcommand{\cwv}{0.22}
\begin{table*}\sffamily
\centering
\begin{tabularx}{\textwidth}{ P{\cwv\textwidth} P{\cwv\textwidth}  P{\cwv\textwidth}  P{\cwv\textwidth}} 
\toprule
 Waveform & Spectrogram &  Waveform & Spectrogram \\ 
\midrule
\includegraphics[width=\cwv\textwidth]{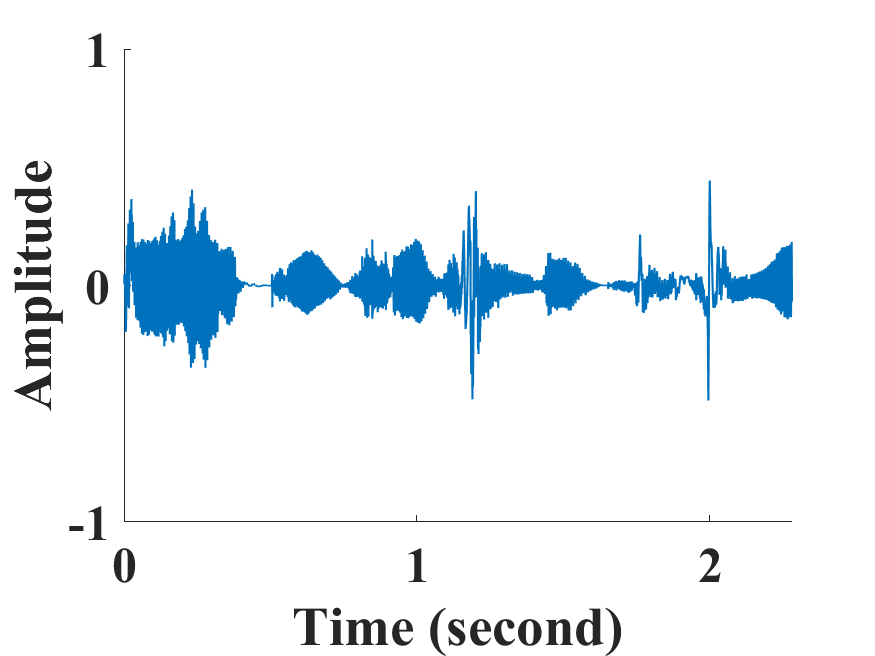} & \includegraphics[width=\cwv\textwidth]{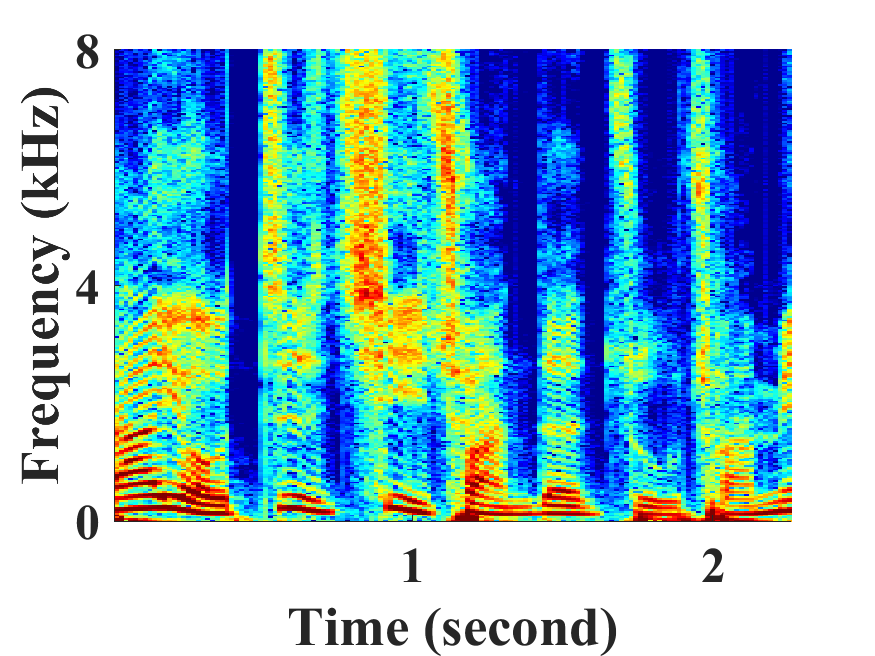} &
\includegraphics[width=\cwv\textwidth]{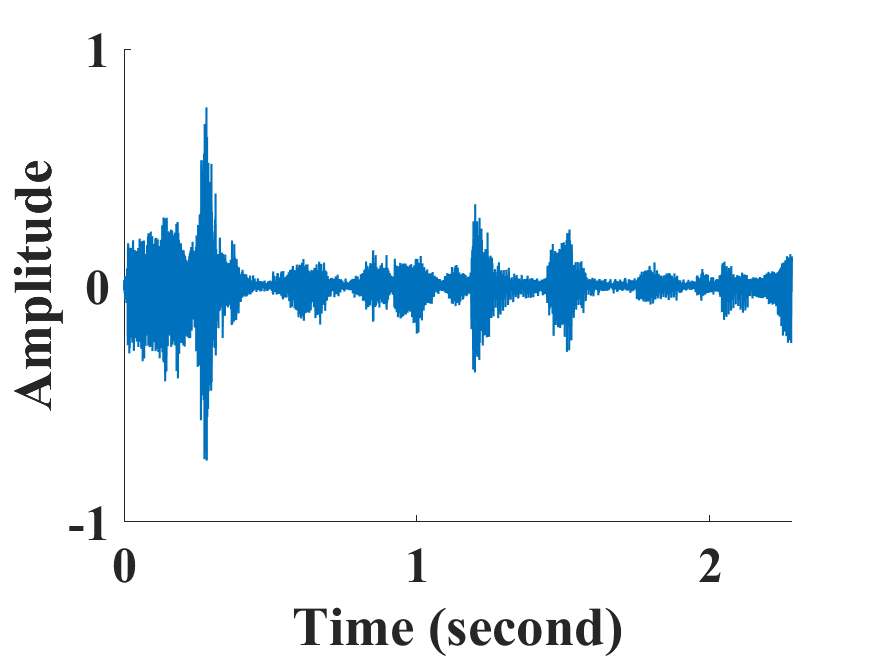} &
\includegraphics[width=\cwv\textwidth]{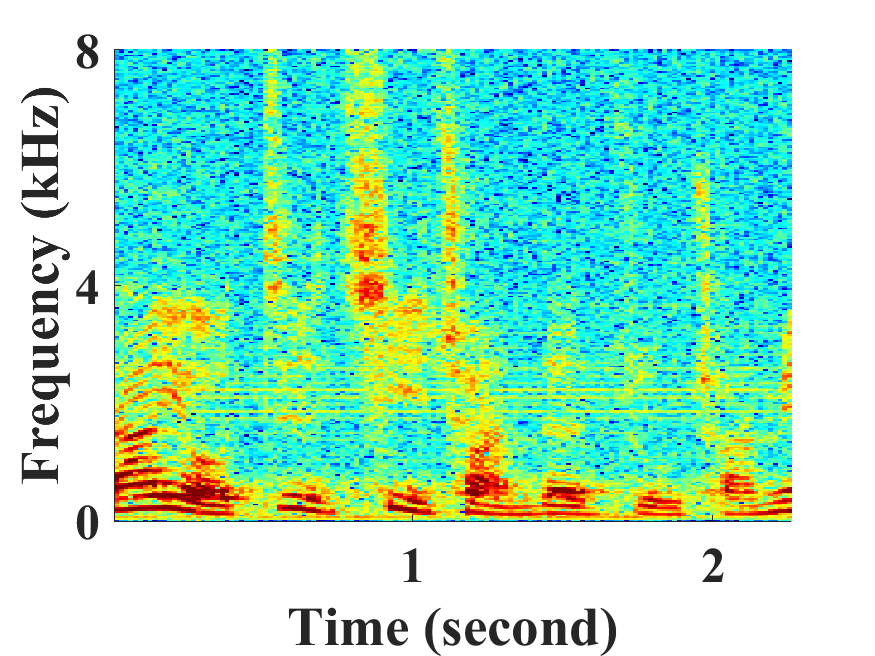} \\ 
\hline
\multicolumn{2}{c}{(a) Near-field microphone} &
\multicolumn{2}{c}{(b) Far-field microphone (II)} \\
\hline
\includegraphics[width=\cwv\textwidth]{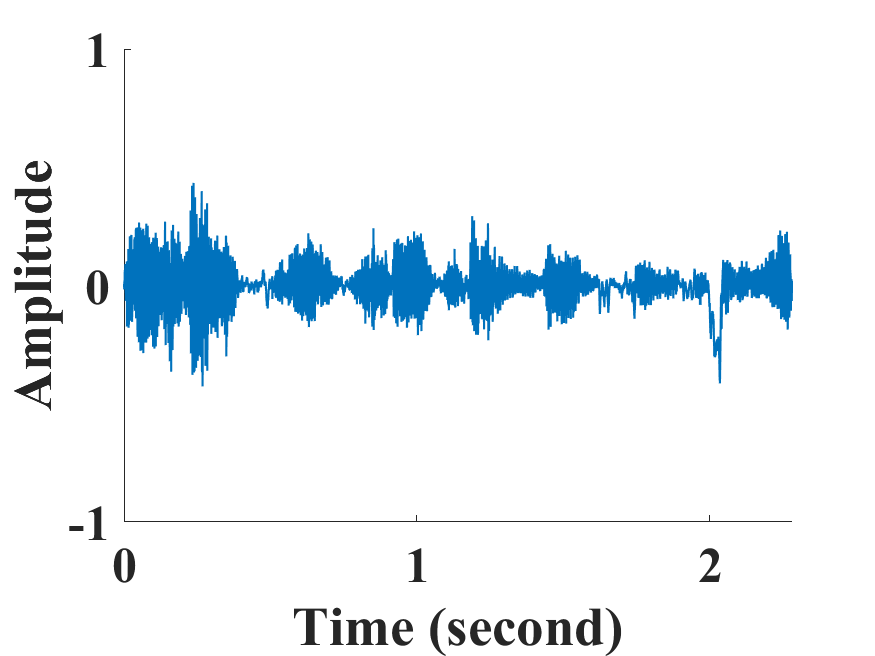} & \includegraphics[width=\cwv\textwidth]{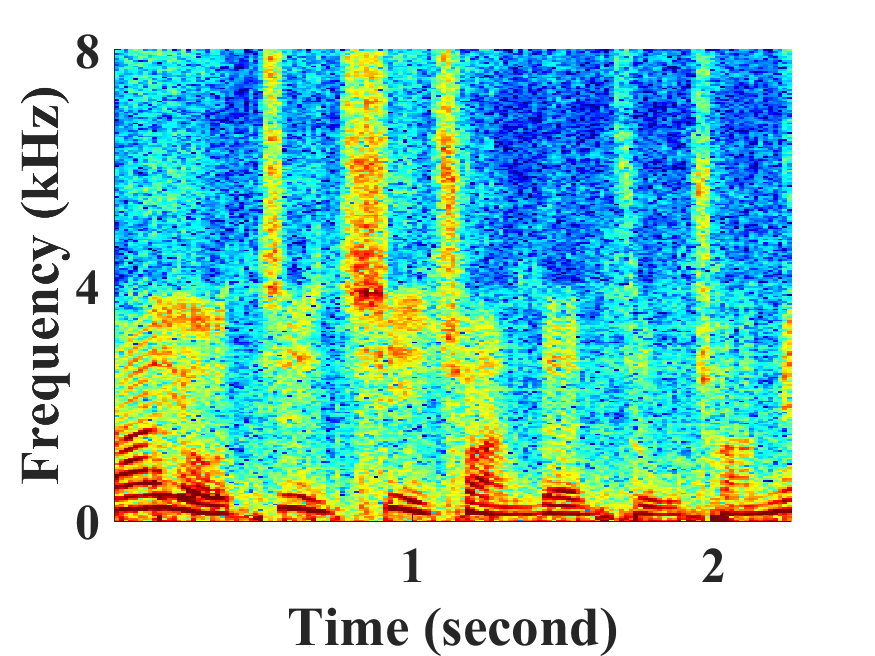} &
\includegraphics[width=\cwv\textwidth]{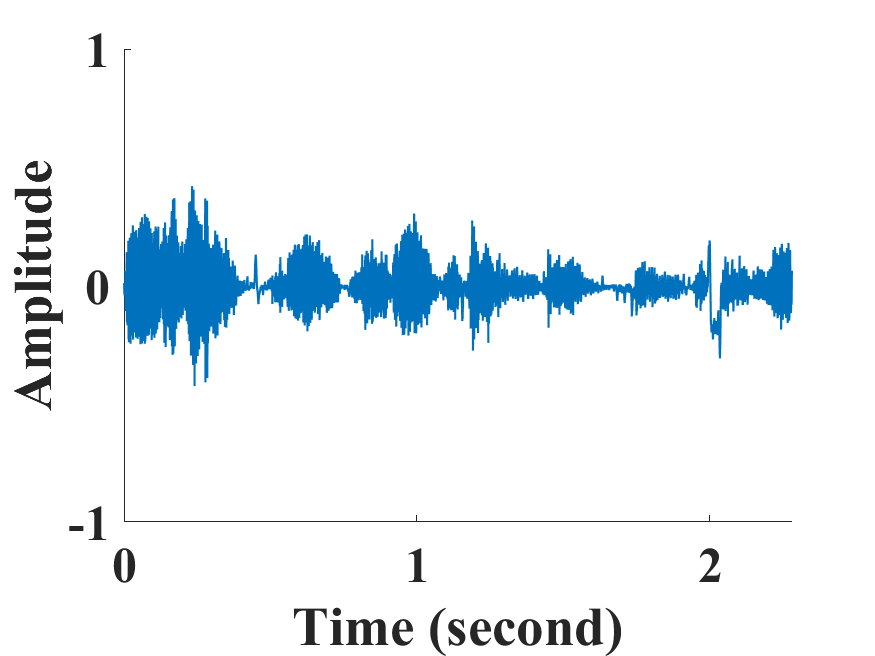} &
\includegraphics[width=\cwv\textwidth]{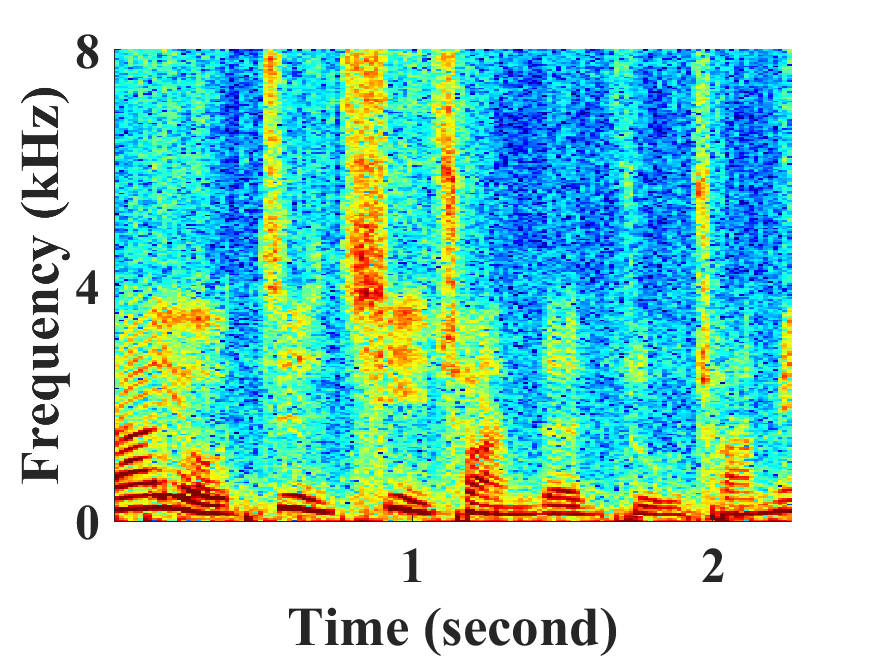} \\ 
\hline
\multicolumn{2}{c}{(c) rSDFCN (Single-channel II)} &
\multicolumn{2}{c}{(d) rSDFCN (Multichannel: I, II, III, IV, V)} \\
\hline
\includegraphics[width=\cwv\textwidth]{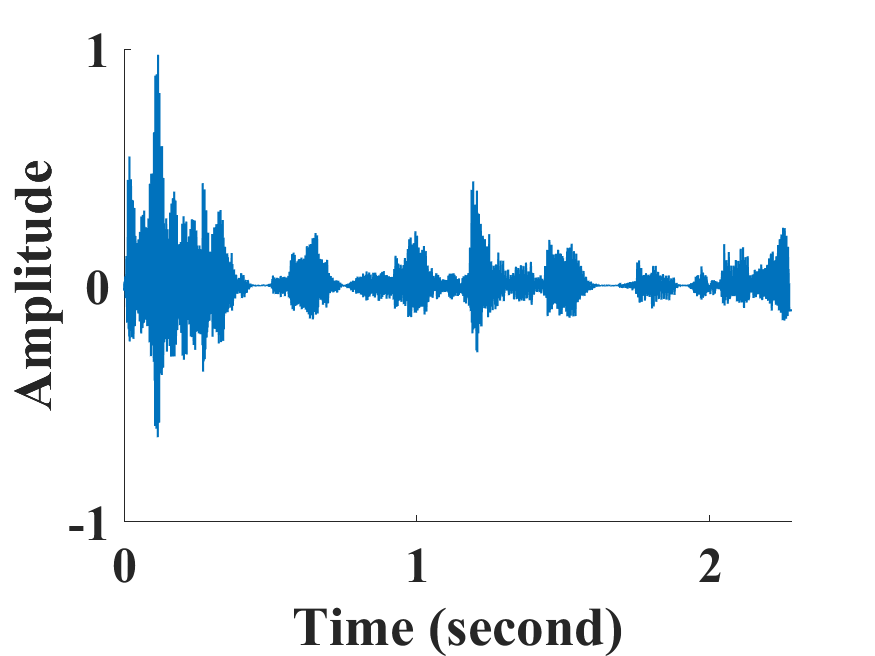} & \includegraphics[width=\cwv\textwidth]{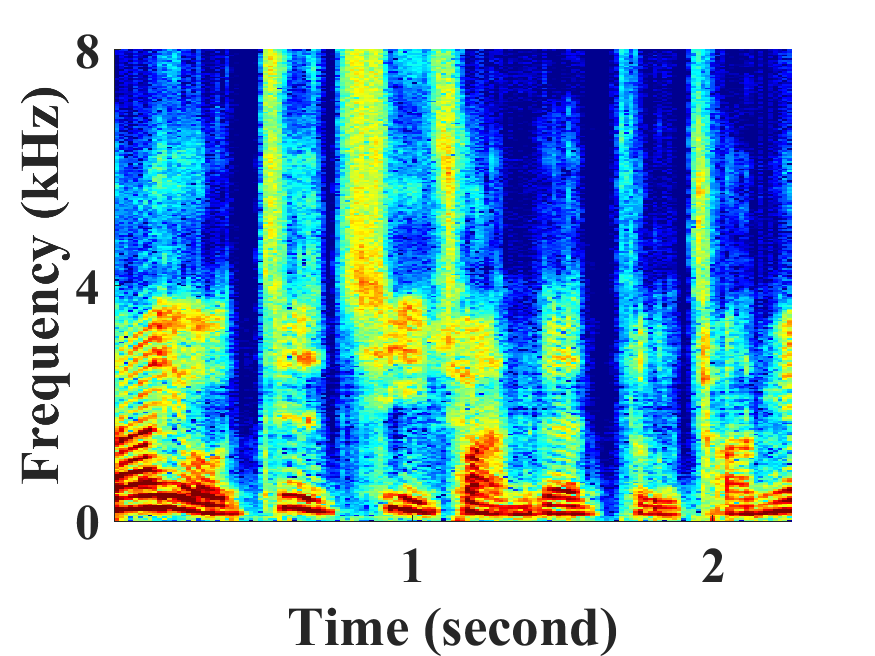} &
\includegraphics[width=\cwv\textwidth]{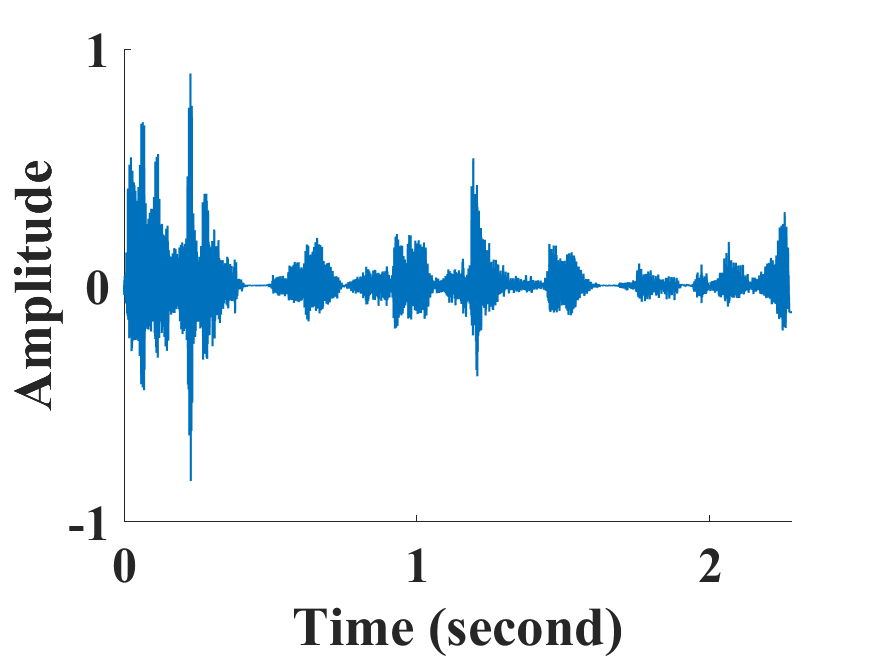} &
\includegraphics[width=\cwv\textwidth]{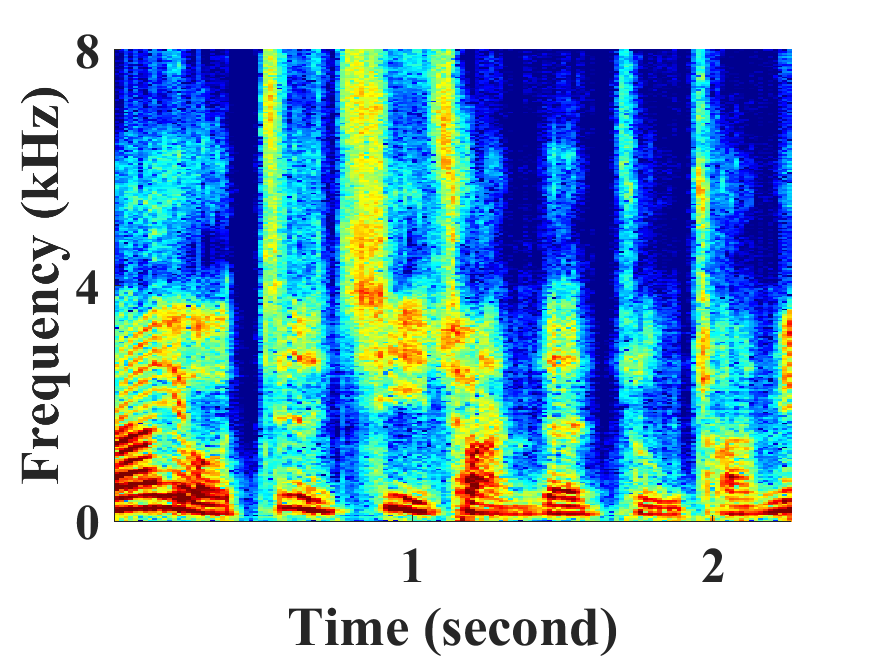} \\ 
\hline
\multicolumn{2}{c}{(e) DDAE (Single-channel II)} &
\multicolumn{2}{c}{(f) DDAE (Multichannel: I, II, III, IV, V)} \\
\bottomrule 
\end{tabularx}
\captionof{figure}{Waveforms and spectrograms of an example utterance in the DM-SE task: (a) speech recorded by near-field microphone; (b) speech recorded by second far-field microphone (channel II); (c) and (e) enhanced speech by rSDFCN and DDAE with single-channel in-put; (d) and (f) enhanced speech by rSDFCN and DDAE with five channels of input.}
\label{fig:f12}
\end{table*}

\begin{figure}
  \centering
  \includegraphics[width=0.48\textwidth]{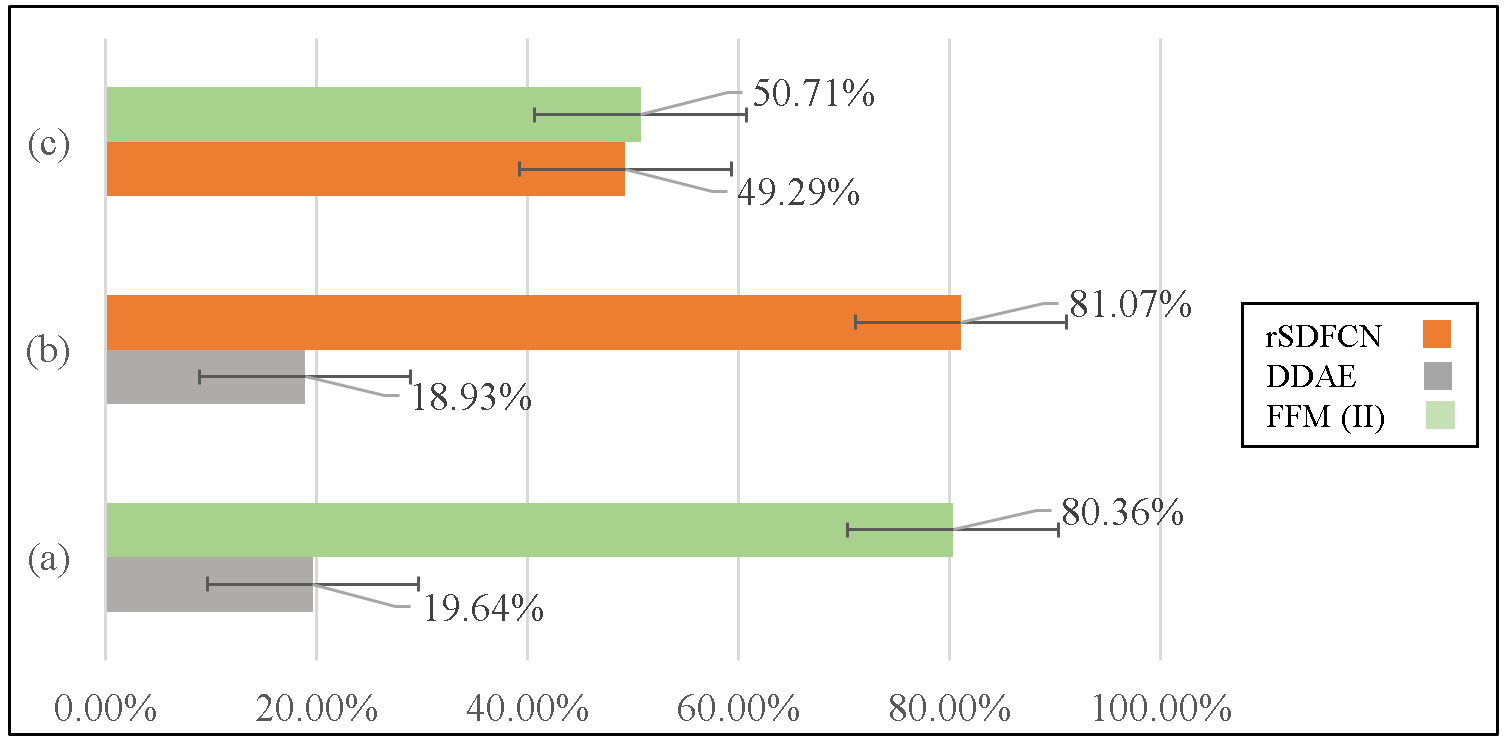}
  \caption{Results of the AB preference test (with 95\% confidence intervals) on speech quality compared between rSDFCN, FFM(II) and DDAE for the DM-SE task.}
  \label{fig:s2}
\end{figure}

We also conducted listening tests on the enhanced speeches by rSDFCN, DDAE and the recorded speech by the second far-field microphone, termed FFM(II) (the channel II in Fig. \ref{fig:f11}, which achieved the highest PESQ score, as shown in Table IV). The results are shown in Fig. \ref{fig:s2}.  From the figure, we note that DDAE cannot improve the speech quality effectively. A possible reason is that the distortions caused by distance does not affect the speech quality too much. Thus, although the DDAE approach can recover missing speech signal components, it may generate distortions and accordingly deteriorate the speech quality. In the meanwhile, we note that the rSDFCN can yield higher speech quality scores than the DDAE, confirming that rSDFCN is superior to DDAE in terms of subjective listening evaluations. Finally, we note that the rSDFCN enhanced speech and the one recorded by the second far-field microphone give comparable listening preference scores (50.71\% versus 49.29\%).

\begin{figure}
  \centering
  \includegraphics[width=0.48\textwidth]{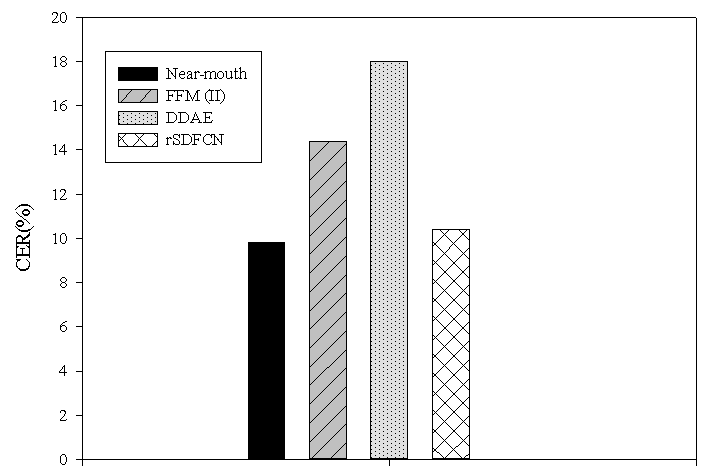}
  \caption{ASR results achieved by different SE models for DM-SE task.}
  \label{fig:f13}
\end{figure}

The recognition results using Google ASR are shown in Fig. \ref{fig:f13}. We report the performance of the speech recorded by the near-field microphone (as the upper-bound) and the second far-field microphone, namely, FFM(II) (channel II in Fig. \ref{fig:f11}, which achieved the best ASR results in our experiments) and the enhanced speech by DDAE and rSDFCN; the corresponding CERs are 9.8\%, 14.4\%, 18.0\%, and 10.4\%, respectively. From the CERs in Fig. \ref{fig:f13}, we first note a clear drop in ASR performance from near-field microphone speech to far-field microphone speech. Next, we note that the CER of the rSDFCN enhanced speech (10.4\%) is much lower than that of the far-field microphone speech (14.0\%) and close to that of the near-field microphone speech (9.8\%). More specifically, the rSDFCN multichannel SE system reduced the CER by 27.8\% (from 14.4\% to 10.4\%) compared to the unenhanced single-channel far-field microphone speech. Comparing the results in Figs. \ref{fig:s2} and \ref{fig:f13} and Table IV, we note that rSDFCN outperforms DDAE in terms of PESQ, STOI, subjective preference test scores, and ASR performance, confirming the effectiveness of the proposed rSDFCN over the conventional DDAE approach for the DM-SE task.

\begin{figure*}[h]
\centering
\subfloat[][]{
   \includegraphics[width=0.4\linewidth,height=4.6cm]{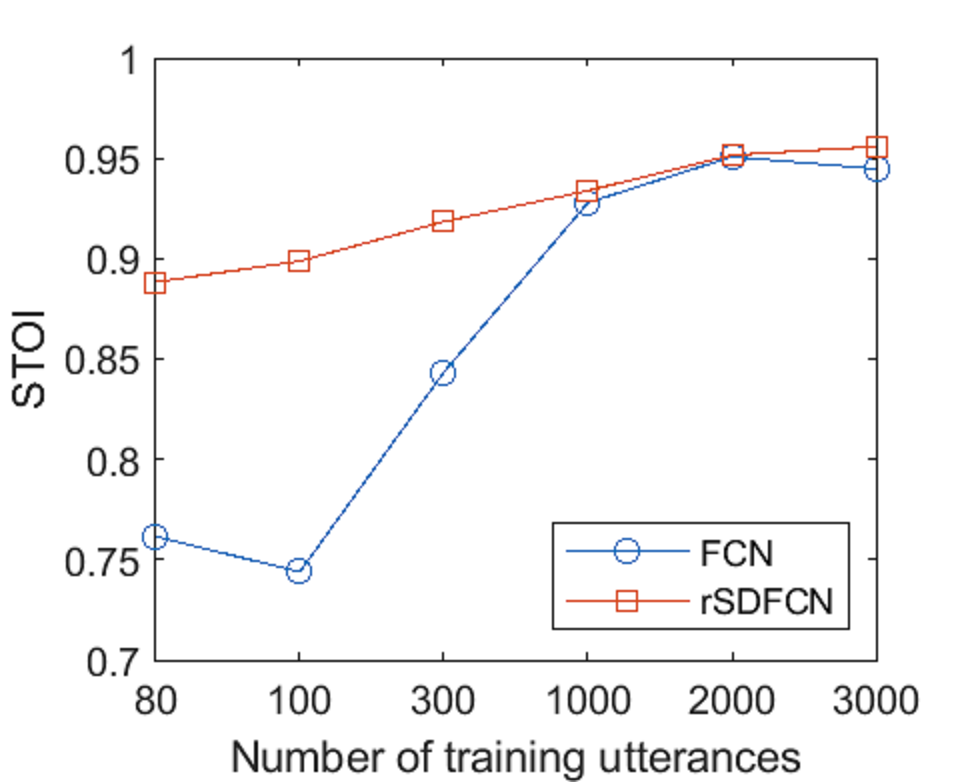}
 }
\subfloat[][]{
   \includegraphics[width=0.4\linewidth,height=4.6cm]{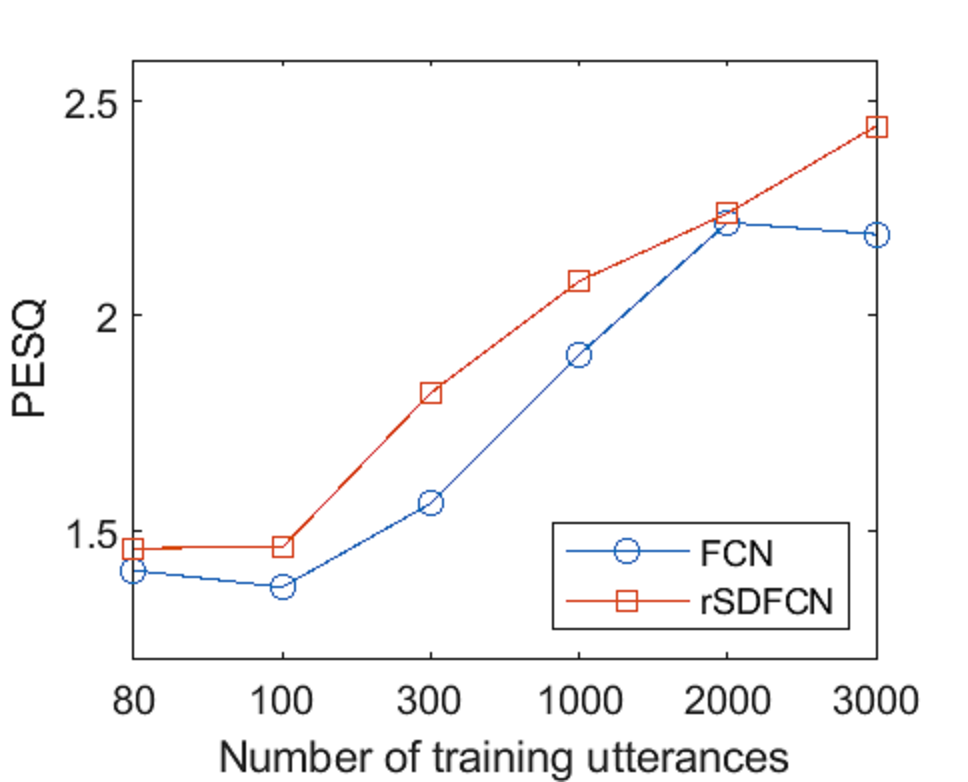}
 }
\caption{Average (a) STOI and (b) PESQ scores of FCN and rSDFCN with different numbers of training utterances in the CHiME-3 dataset.}
\label{fig:chime_set_size}
\end{figure*}

\textcolor{ColorVariable1}{
\subsection{Speech Enhancement on the CHiME-3 dataset}
To  further validate  the  effectiveness  of  using  multichannel inputs  for  SE, we also tested our rSDFCN system on the CHiME-3 dataset \cite{barker2015third}. As documented, the clean (reference) speech in the CHiME-3 training set was directly copied from the WSJ0 corpus, while the reference speech in the CHiME-3 testing set was generated from the booth recording. It is not fair to use the booth-recorded data as the reference to compute the STOI and PESQ scores of the enhanced speech. We tested our rSDFCN system on the simulated speech data of the CHiME-3 dataset. The simulated data is built by mixing clean speeches of the Wall Street Journal (WSJ0) corpus \cite{garofalo2007csr} with four different real  background noises: bus (\textbf{BUS}), cafeteria (\textbf{CAF}), pedestrian zone (\textbf{PED}) and street (\textbf{STR}). All the clean speeches and the noises are recorded by a 6-microphone array on a tablet. The total simulated set contains 7138 utterances, including 1728 of \textbf{BUS}, 1794 of \textbf{CAF}, 1765 of \textbf{PED}, and 1851 of \textbf{STR}. The goal is to use recorded six-channel noisy speeches as the input to generate enhanced speech. In our experiments, {\color{ColorVariable3} we trimmed all utterances to speech segments, each containing 36,500 sample points (around 2.28 seconds)}. Because the CHiME-3 dataset is far larger than the two datasets used in previous experiments, we also conducted experiments to explore the enhancement performance with respect to different numbers of training utterances. Note that in this experiment, we trained our model on utterances of \textbf{PED}, \textbf{STR} and \textbf{CAF}, and tested them on \textbf{BUS} because \textbf{BUS} was the most difficult for rSDFCN to archieve improvements over DDAE and FCN in our preliminary experiments. Fig. \ref{fig:chime_set_size} shows the STOI and PESQ scores of FCN and rSDFCN with respect to different numbers of training utterances. From Fig. \ref{fig:chime_set_size}(a), we can see that rSDFCN, which contains the dilated and Sinc convolutional layers, achieves much higher STOI scores than FCN when the number of training utterances is limited. This implies that the benefits of the dilated and Sinc convolutional layers are more significant when the training set is small.}

\begin{table*}
\color{ColorVariable2}
\centering
\caption{ Average STOI/PESQ Scores of rSDFCN and DDAE evalauated on the CHiME-3 dataset.}
\begin{tabular}{|P{3cm}|| P{1.8cm}| P{1.8cm} |P{1.8cm} || P{1.8cm}| P{1.8cm} |P{1.8cm} |}
\hline
AVG. STOI/PESQ      & \multicolumn{3}{c|}{STOI}           & \multicolumn{3}{c|}{PESQ}               \\ \hline
Input Microphone(s) & Unenhanced & DDAE  & rSDFCN         & Unenhanced     & DDAE  & rSDFCN         \\ \hline
I                   & 0.847      & 0.825 & \textbf{0.878} & 1.208          & 1.478 & \textbf{1.592} \\ \hline
II                  & 0.863      & 0.826 & \textbf{0.880} & 1.244          & 1.465 & \textbf{1.604} \\ \hline
III                 & 0.844      & 0.828 & \textbf{0.880} & 1.198          & 1.466 & \textbf{1.620} \\ \hline
IV                  & 0.884      & 0.824 & \textbf{0.879} & 1.308          & 1.462 & \textbf{1.614} \\ \hline
V                   & 0.893      & 0.827 & \textbf{0.876} & 1.337          & 1.469 & \textbf{1.598} \\ \hline
VI                  & 0.833      & 0.825 & \textbf{0.880} & 1.185          & 1.466 & \textbf{1.628} \\ \hline
I– VI               &            & 0.853 & \textbf{0.937} &                & 1.621 & \textbf{2.145} \\ \hline

\end{tabular}
\label{table:chime_all}
\end{table*}

\textcolor{ColorVariable1}{
Next, Table \ref{table:chime_all} shows the average STOI and
PESQ scores of rSDFCN and DDAE with single-channel and multichannel inputs. Since there are four types of the background noises, we set utterances with one type of noise as the test set and use all utterances with the other three types of noises as the training sets in turn. This leave-one-out training and testing procedure repeated four times, and the average STOI and PESQ scores from the four sets of results were reported in Table \ref{table:chime_all}. 
Similar to the trends in the previous two datasets, Table \ref{table:chime_all} shows that the scores of multichannel-based rSDFCN are much higher than those of DDAE and single-channel-based rSDFCN.}

\section{CONCLUSION}
In this paper, we proposed the SDFCN waveform-mapping-based multichannel SE system and an extended version, rSDFCN, to further improve the performance. We tested the proposed SE systems on three multichannel SE tasks: IEM-SE, DM-SE \textcolor{ColorVariable1}{and CHiME-3}. The experimental results for the three tasks confirmed the effectiveness of the proposed systems in achieving higher STOI and PESQ scores, as well as providing higher subjective listening scores and improved ASR performance. Meanwhile, the proposed waveform-based rSDFCN SE system outperformed the spectral-mapping-based DDAE SE system, which confirms that phase information is important for multichannel SE.\par
To the best of our knowledge, this study is one of the first works that adopt the concept of waveform mapping based on neural network models to enhance multichannel speech signals. In this work, both IEM-SE and DM-SE tasks simulated a “virtual” high-performance and near-field microphone to overcome the distortion caused by channel effects and spatial fading, and to attain improved speech quality (PESQ), speech intelligibility (STOI), subjective listening scores, and ASR performance. \textcolor{ColorVariable1}{The proposed system also shows promising performance on the standardized CHiME-3 dataset.} Please note that different from the beamforming methods that require spatial and time-delay information, this study investigates the scenario where the speech signals are recorded by multiple microphones simultaneously. In the future, we will extend the proposed systems to multichannel tasks where multiple distortion factors including noise, interference, and reverberation are involved. Meanwhile, we will explore the possibility of combining the advantages of waveform-mapping and spectral-mapping-based multichannel SE methods to further improve our current systems.

\section*{Acknowledgment}
The authors would like to thank the financial support pro-vided by Ministry of Science and Technology, Taiwan (106-2221-E-001-017-MY2 and 107-2221-E-001-012-MY2).

% Can use something like this to put references on a page
% by themselves when using endfloat and the captionsoff option.
\ifCLASSOPTIONcaptionsoff
  \newpage
\fi

\begin{IEEEbiography}
[{\includegraphics[width=1in,height=1.25in,clip,keepaspectratio]{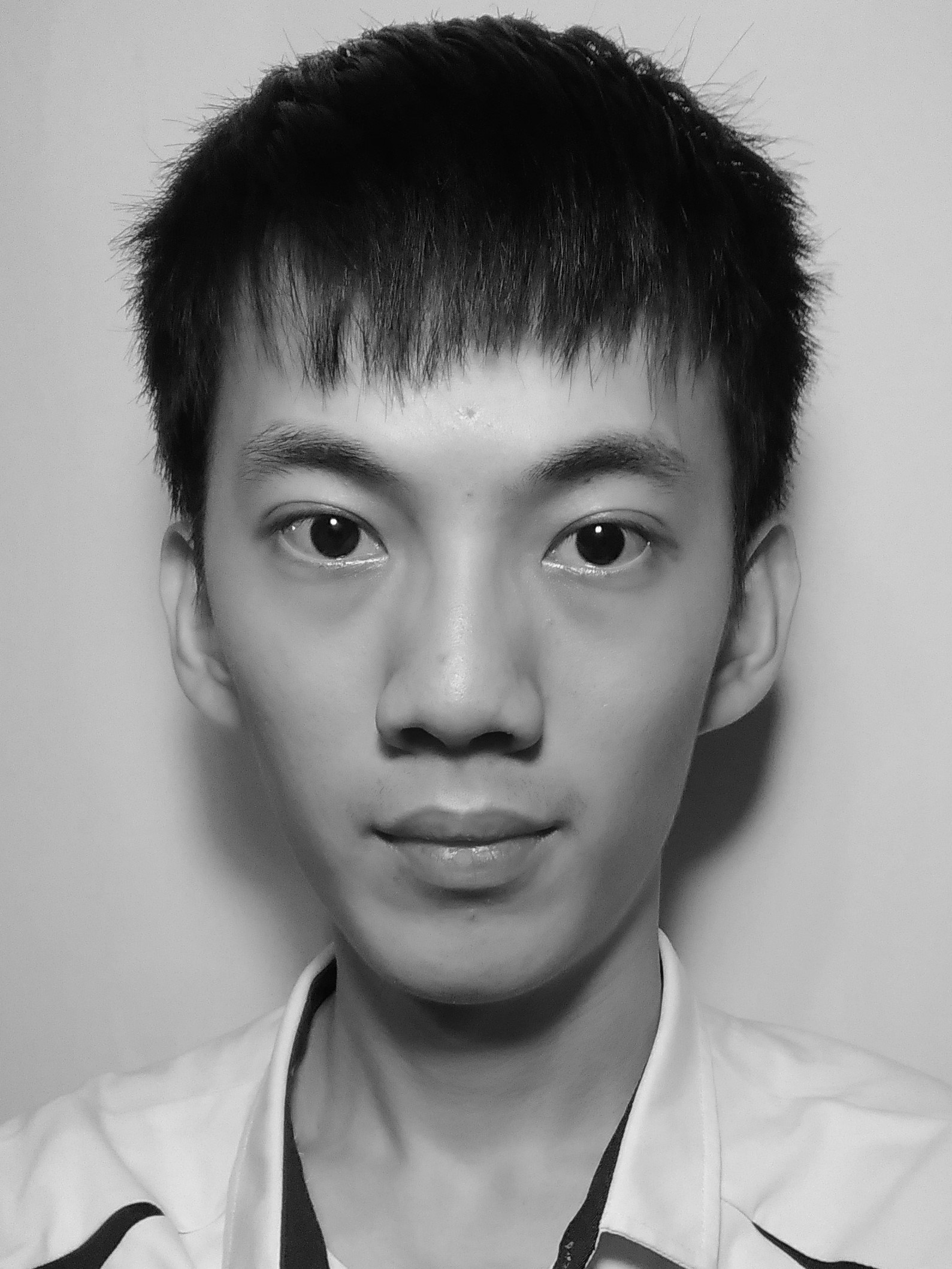}}]%
{Chang-Le Liu} is currently working toward the B.S. degree in electrical engineering in National Taiwan University, Taipei, Taiwan, from 2016. 
He was a intern as a research assistant with the Research Center for Information Technology Innovation, Academia Sinica, Taipei, Taiwan, and was involved in research in speech enhancement. His current research topic includes audio and image signal processing.
\end{IEEEbiography}
\begin{IEEEbiography}
[{\includegraphics[width=1in,height=1.25in,clip,keepaspectratio]{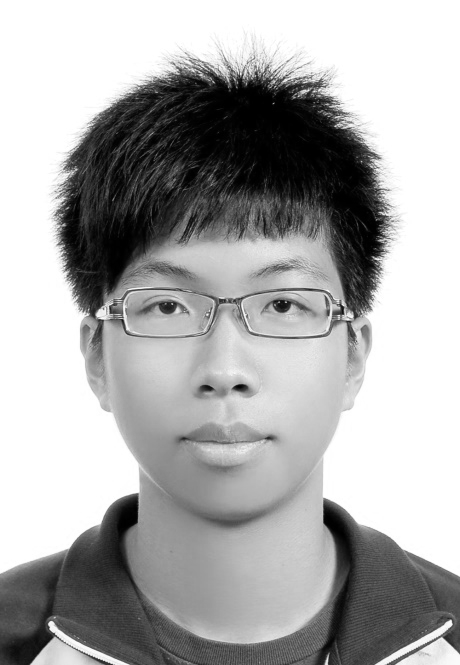}}]%
{Sze-Wei Fu} received the B.S. and M.S. degrees in Department of Engineering Science and Ocean Engineering and Graduate Institute of Communication Engineering from National Taiwan University, Taipei, Taiwan, in 2012 and 2014, respectively. He is currently pursuing the Ph.D. degree with the Department of Computer Science and Information Engineering, National Taiwan University, Taipei and he is also a Research Assistant in the Research Center for Information Technology Innovation, Academia Sinica, Taiwan. His research interests include speech processing, speech enhancement, machine learning and deep learning.
\end{IEEEbiography}
\begin{IEEEbiography}
[{\includegraphics[width=1in,height=1.25in,clip,keepaspectratio]{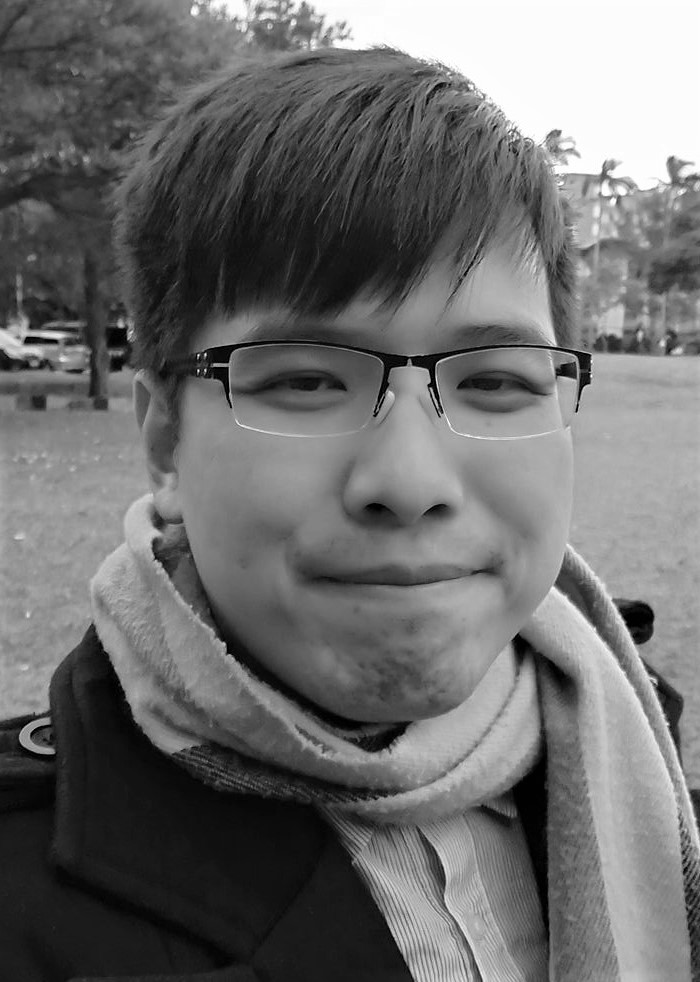}}]%
{You-Jin Li} received the B.S. degree in Department of electronic engineering from the National Ilan University, Ilan, Taiwain, in 2014, and the M.S. degree in Department of electrical engineering with communications from the National Ilan University, Ilan, Taiwain, in 2016. He is currently pursuing the Ph.D. degree with the Graduate Institute of Communication Engineering, National Taiwan University, Taipei, Taiwain. His research interests cover signal processing, speech enhancement, beamforming, deep learning, and multi-channel compression.
\end{IEEEbiography}
\begin{IEEEbiography}
[{\includegraphics[width=1in,height=1.25in,clip,keepaspectratio]{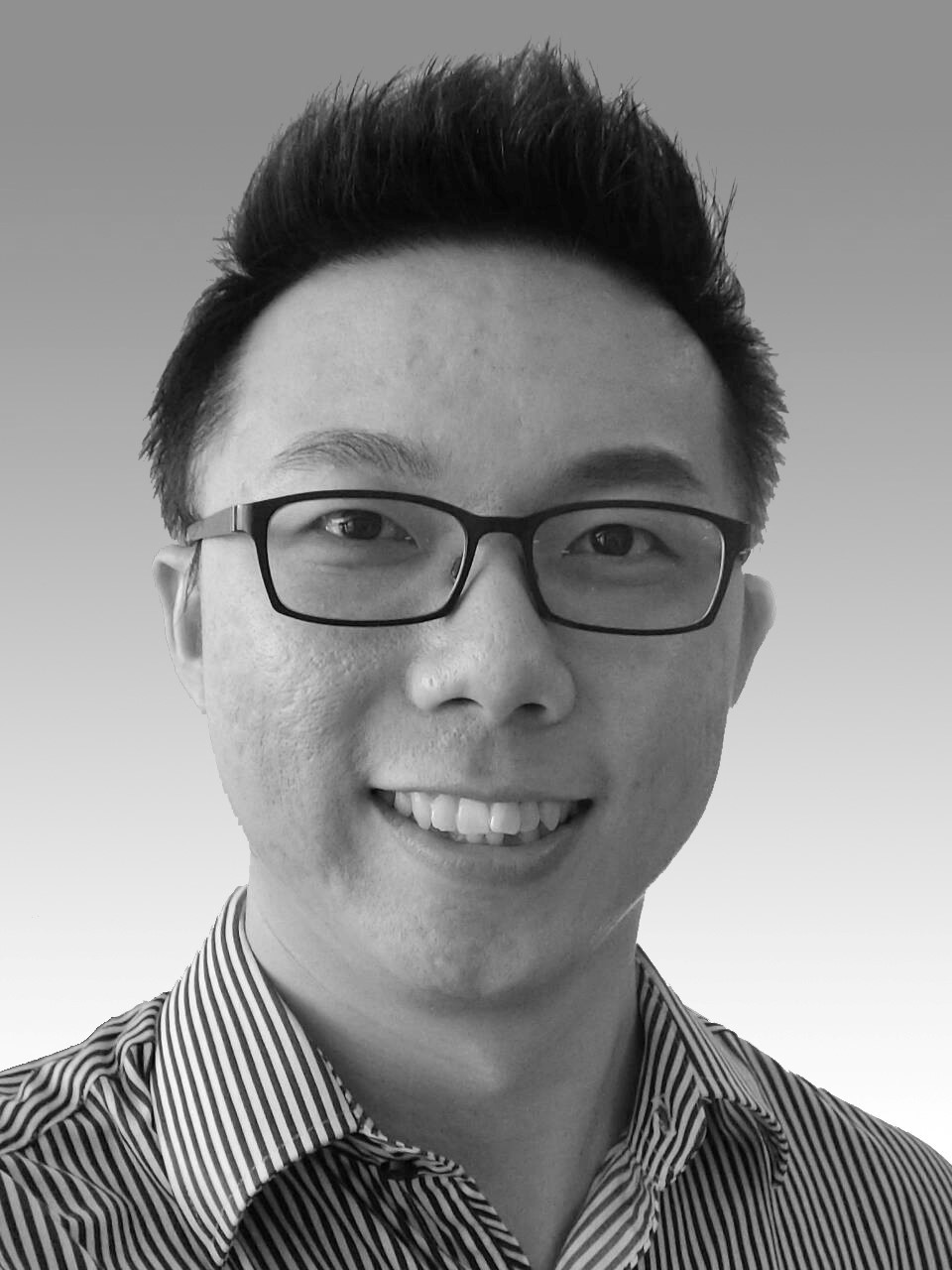}}]%
{Jen-Wei Hunag} received the BS and PhD degree in electrical engineering from National Taiwan University, Taiwan in 2002 and 2009 respectively. He was a visiting scholar in IBM Almaden Research Center from 2008 to 2009, an assistant professor in Yuan Ze University from 2009 to 2012, and a visiting scholar in University of Chicago in 2016. He is now an associate professor in the Department of Electrical Engineering, National Cheng Kung University, Taiwan. He serves as the Director of Taiwanese Association of Artificial Intelligence. He was the committee member of IEEE CIS Member Activities committee from 2017 to 2018 and the secretary of IEEE Tainan Section CIS Chapter from 2015 to 2017. His major research topics are Data Mining, Machine Learning and Artificial Intelligence. Among these, social network analysis, spatial-temporal data mining, text mining and multimedia information retrieval are his special interests. In addition, some of his research are on FinTech and bioinformatics.
\end{IEEEbiography}
\begin{IEEEbiography}
[{\includegraphics[width=1in,height=1.25in,clip,keepaspectratio]{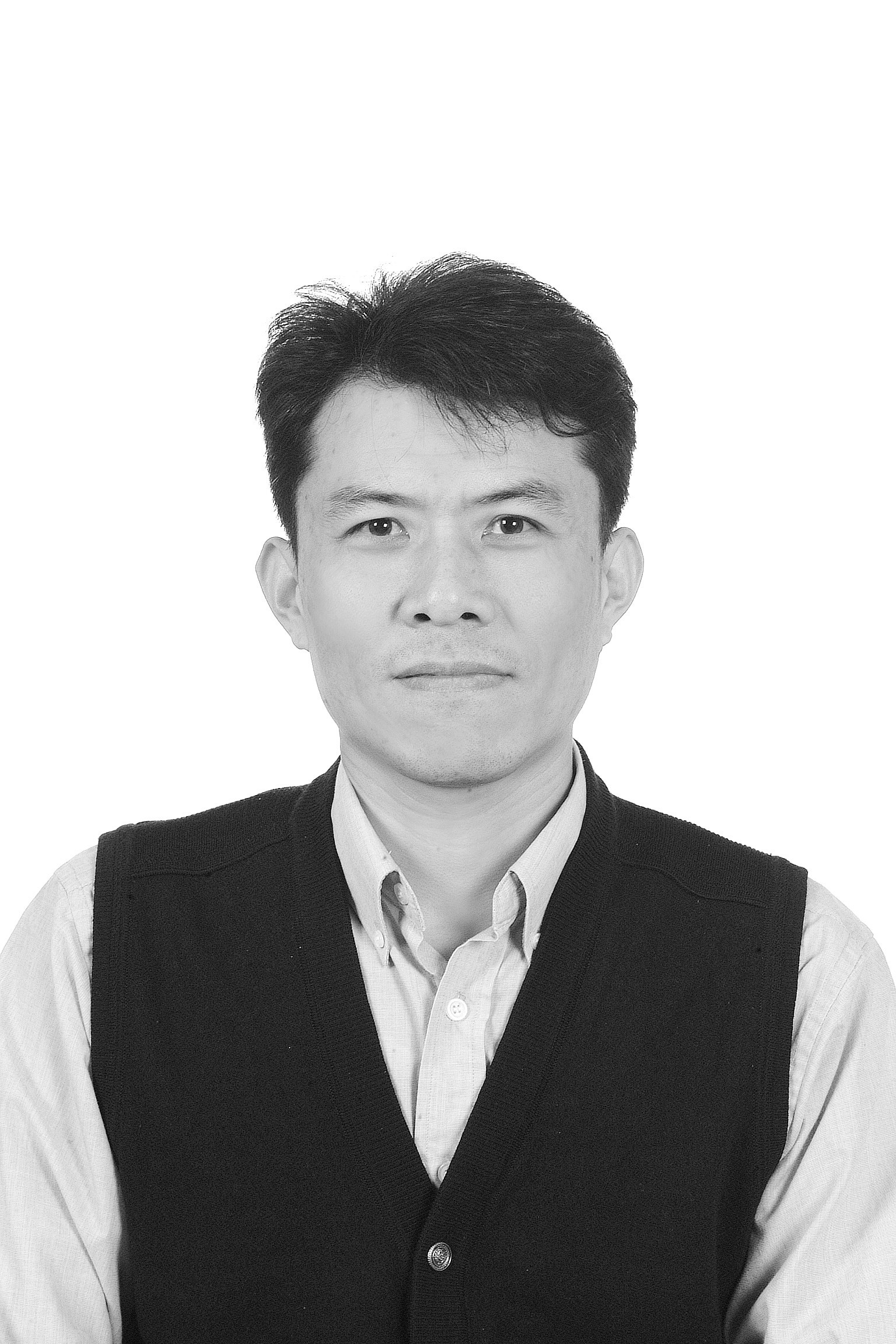}}]%
{Hsin-Min Wang} (S’92–M’95–SM’04) received the B.S. and Ph.D. degrees in electrical engineering from National Taiwan University, Taipei, Taiwan, in 1989 and 1995, respectively. In October 1995, he joined the Institute of Information Science, Academia Sinica, Taipei, Taiwan, where he is currently a Research Fellow. He also holds a joint appointment as a Professor in the Department of Computer Science and Information Engineering at National Cheng Kung University. He currently serves an Editorial Board Member of IEEE/ACM Transactions on Audio, Speech and Language Processing and APSIPA Transactions on Signal and Information Processing. His major research interests include spoken language processing, natural language processing, multimedia information retrieval, machine learning and pattern recognition. He was a General Co-Chair of ISCSLP2016 and ISCSLP2018 and a Technical Co-Chair of ISCSLP2010, O-COCOSDA2011, APSIPAASC2013, ISMIR2014, and ASRU2019. He received the Chinese Institute of Engineers Technical Paper Award in 1995 and the ACM Multimedia Grand Challenge First Prize in 2012. He was an APSIPA distinguished lecturer for 2014–2015. He is a member of the International Speech Communication Association and ACM.
\end{IEEEbiography}
\begin{IEEEbiography}
[{\includegraphics[width=1in,height=1.25in,clip,keepaspectratio]{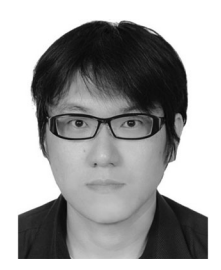}}]%
{Yu Tsao} (M’09) received the B.S. and M.S. degrees in electrical engineering from National Taiwan University, Taipei, Taiwan, in 1999 and 2001, respectively, and the Ph.D. degree in electrical and computer engineering from the Georgia Institute of Technology, Atlanta, GA, USA, in 2008. From 2009 to 2011, he was a Researcher with the National Institute of Information and Communications Technology, Tokyo, Japan, where he engaged in research and product development in automatic speech recognition for multilingual speech-to-speech translation. He is currently an Associate Research Fellow with the Research Center for Information Technology Innovation, Academia Sinica, Taipei, Taiwan. His research interests include speech and speaker recognition, acoustic and language modeling, audio coding, and bio-signal processing. He is currently an Associate Editor of the IEEE/ACM Transactions on Audio, Speech, and Language Processing and IEICE transactions on Information and Systems. Dr. Tsao received the Academia Sinica Career Development Award in 2017, National Innovation Award in 2018 and 2019, and Outstanding Elite Award, Chung Hwa Rotary Educational Foundation 2019-2020.  
\end{IEEEbiography}

\vfill


\begin{thebibliography}{10}

\providecommand{\url}[1]{#1}
\csname url@samestyle\endcsname
\providecommand{\newblock}{\relax}
\providecommand{\bibinfo}[2]{#2}
\providecommand{\BIBentrySTDinterwordspacing}{\spaceskip=0pt\relax}
\providecommand{\BIBentryALTinterwordstretchfactor}{4}
\providecommand{\BIBentryALTinterwordspacing}{\spaceskip=\fontdimen2\font plus
\BIBentryALTinterwordstretchfactor\fontdimen3\font minus
  \fontdimen4\font\relax}
\providecommand{\BIBforeignlanguage}[2]{{%
\expandafter\ifx\csname l@#1\endcsname\relax
\typeout{** WARNING: IEEEtran.bst: No hyphenation pattern has been}%
\typeout{** loaded for the language `#1'. Using the pattern for}%
\typeout{** the default language instead.}%
\else
\language=\csname l@#1\endcsname
\fi
#2}}
\providecommand{\BIBdecl}{\relax}
\BIBdecl

\bibitem{Zhao2018Convolutional}
\BIBentryALTinterwordspacing
Z.~Zhao, H.~Liu, and T.~Fingscheidt, ``Convolutional neural networks to enhance
  coded speech,'' \emph{arXiv:1806.09411 [cs, eess]}, 6 2018, arXiv:
  1806.09411. [Online]. Available: \url{http://arxiv.org/abs/1806.09411}
\BIBentrySTDinterwordspacing

\bibitem{Li2011Comparative}
J.~Li, L.~Yang, J.~Zhang, Y.~Yan, Y.~Hu, M.~Akagi, and P.~C. Loizou,
  ``\BIBforeignlanguage{en}{Comparative intelligibility investigation of
  single-channel noise-reduction algorithms for chinese, japanese, and
  english},'' \emph{\BIBforeignlanguage{en}{The Journal of the Acoustical
  Society of America}}, vol. 129, no.~5, pp. 3291--3301, 5 2011.

\bibitem{Chen2016Large-scale}
J.~Chen, Y.~Wang, S.~E. Yoho, D.~Wang, and E.~W. Healy,
  ``\BIBforeignlanguage{eng}{Large-scale training to increase speech
  intelligibility for hearing-impaired listeners in novel noises},''
  \emph{\BIBforeignlanguage{eng}{The Journal of the Acoustical Society of
  America}}, vol. 139, no.~5, p. 2604, 2016, pMID: 27250154 PMCID: PMC5392064.

\bibitem{Lai2017Deep}
Y.-H. Lai, F.~Chen, S.-S. Wang, X.~Lu, Y.~Tsao, and C.-H. Lee,
  ``\BIBforeignlanguage{en}{A deep denoising autoencoder approach to improving
  the intelligibility of vocoded speech in cochlear implant simulation},''
  \emph{\BIBforeignlanguage{en}{IEEE Transactions on Biomedical Engineering}},
  vol.~64, no.~7, pp. 1568--1578, 7 2017.

\bibitem{Li2015Robust}
J.~Li, L.~Deng, R.~Haeb-Umbach, and Y.~Gong, \emph{Robust automatic speech
  recognition: a bridge to practical applications}.\hskip 1em plus 0.5em minus
  0.4em\relax Academic Press, 2015.

\bibitem{Boll1979Suppression}
S.~Boll, ``Suppression of acoustic noise in speech using spectral
  subtraction,'' \emph{IEEE Transactions on Acoustics, Speech, and Signal
  Processing}, vol.~27, no.~2, pp. 113--120, 4 1979.

\bibitem{Krishnamoorthi2010auditory-domain}
H.~Krishnamoorthi, A.~Spanias, V.~Berisha, H.~Kwon, and H.~Thornburg, ``An
  auditory-domain based speech enhancement algorithm.''\hskip 1em plus 0.5em
  minus 0.4em\relax 2010 IEEE International Conference on Acoustics, Speech and
  Signal Processing, 3 2010, pp. 4786--4789.

\bibitem{McAulay1980Speech}
R.~McAulay and M.~Malpass, ``Speech enhancement using a soft-decision noise
  suppression filter,'' \emph{IEEE Transactions on Acoustics, Speech, and
  Signal Processing}, vol.~28, no.~2, pp. 137--145, 4 1980.

\bibitem{Ephraim1984Speech}
Y.~Ephraim and D.~Malah, ``Speech enhancement using a minimum-mean square error
  short-time spectral amplitude estimator,'' \emph{IEEE Transactions on
  Acoustics, Speech, and Signal Processing}, vol.~32, no.~6, pp. 1109--1121, 12
  1984.

\bibitem{Loizou2003generalized}
Loizou and P.~C., ``A generalized subspace approach for enhancing speech
  corrupted by colored noise,'' \emph{IEEE Transactions on Speech and Audio
  Processing}, vol.~11, no.~4, pp. 334--341, 7 2003.

\bibitem{Rezayee2001adaptive}
A.~Rezayee and S.~Gazor, ``An adaptive klt approach for speech enhancement,''
  \emph{IEEE Transactions on Speech and Audio Processing}, vol.~9, no.~2, pp.
  87--95, 2 2001.

\bibitem{Vetter1999Single}
R.~Vetter, N.~Virag, P.~Renevey, and J.-M. Vesin, ``Single channel speech
  enhancement using principal component analysis and mdl subspace selection,''
  1999.

\bibitem{Mohammadiha2013Supervised}
N.~Mohammadiha, P.~Smaragdis, and A.~Leijon, ``Supervised and unsupervised
  speech enhancement using nonnegative matrix factorization,'' \emph{IEEE
  Transactions on Audio, Speech, and Language Processing}, vol.~21, no.~10, pp.
  2140--2151, 10 2013, arXiv: 1709.05362.

\bibitem{Wang2016Compressive}
J.~Wang, Y.~Lee, C.~Lin, S.~Wang, C.~Shih, and C.~Wu, ``Compressive
  sensing-based speech enhancement,'' \emph{IEEE/ACM Transactions on Audio,
  Speech, and Language Processing}, vol.~24, no.~11, pp. 2122--2131, 11 2016.

\bibitem{Sigg2010Speech}
C.~D. Sigg, T.~Dikk, and J.~M. Buhmann, ``Speech enhancement with sparse coding
  in learned dictionaries.''\hskip 1em plus 0.5em minus 0.4em\relax 2010 IEEE
  International Conference on Acoustics, Speech and Signal Processing, 3 2010,
  pp. 4758--4761.

\bibitem{Huang2012Singing-voice}
P.~Huang, S.~D. Chen, P.~Smaragdis, and M.~Hasegawa-Johnson, ``Singing-voice
  separation from monaural recordings using robust principal component
  analysis.''\hskip 1em plus 0.5em minus 0.4em\relax 2012 IEEE International
  Conference on Acoustics, Speech and Signal Processing (ICASSP), 3 2012, pp.
  57--60.

\bibitem{Lu2014Ensemble}
X.~Lu, Y.~Tsao, S.~Matsuda, and C.~Hori, ``Ensemble modeling of denoising
  autoencoder for speech spectrum restoration,'' 2014.

\bibitem{Lu2013Speech}
------, ``Speech enhancement based on deep denoising autoencoder,'' 2013.

\bibitem{Kolbak2017Speech}
M.~Kolbæk, Z.~Tan, and J.~Jensen, ``Speech intelligibility potential of
  general and specialized deep neural network based speech enhancement
  systems,'' \emph{IEEE/ACM Transactions on Audio, Speech, and Language
  Processing}, vol.~25, no.~1, pp. 153--167, 1 2017.

\bibitem{LiuExperiments}
D.~Liu, P.~Smaragdis, and M.~Kim, ``\BIBforeignlanguage{en}{Experiments on deep
  learning for speech denoising},'' p.~5.

\bibitem{Xu2014Experimental}
Y.~Xu, J.~Du, L.~Dai, and C.~Lee, ``An experimental study on speech enhancement
  based on deep neural networks,'' \emph{IEEE Signal Processing Letters},
  vol.~21, no.~1, pp. 65--68, 1 2014.

\bibitem{Xu2015Regression}
------, ``A regression approach to speech enhancement based on deep neural
  networks,'' \emph{IEEE/ACM Transactions on Audio, Speech, and Language
  Processing}, vol.~23, no.~1, pp. 7--19, 1 2015.

\bibitem{Campolucci1999On-line}
P.~Campolucci, A.~Uncini, F.~Piazza, and B.~D. Rao, ``On-line learning
  algorithms for locally recurrent neural networks,'' \emph{IEEE Transactions
  on Neural Networks}, vol.~10, no.~2, pp. 253--271, 3 1999.

\bibitem{Weninger2014Single-channel}
F.~Weninger, F.~Eyben, and B.~Schuller, ``Single-channel speech separation with
  memory-enhanced recurrent neural networks.''\hskip 1em plus 0.5em minus
  0.4em\relax 2014 IEEE International Conference on Acoustics, Speech and
  Signal Processing (ICASSP), 5 2014, pp. 3709--3713.

\bibitem{Fu2017Complex}
\BIBentryALTinterwordspacing
S.-W. Fu, T.-y. Hu, Y.~Tsao, and X.~Lu, ``Complex spectrogram enhancement by
  convolutional neural network with multi-metrics learning,''
  \emph{arXiv:1704.08504 [cs, stat]}, 4 2017, arXiv: 1704.08504. [Online].
  Available: \url{http://arxiv.org/abs/1704.08504}
\BIBentrySTDinterwordspacing

\bibitem{Fu2016SNR-Aware}
\BIBentryALTinterwordspacing
S.-W. Fu, Y.~Tsao, and X.~Lu, ``\BIBforeignlanguage{en}{Snr-aware convolutional
  neural network modeling for speech enhancement}.''\hskip 1em plus 0.5em minus
  0.4em\relax Interspeech 2016, 9 2016, pp. 3768--3772, [Online; accessed
  2018-10-26]. [Online]. Available:
  \url{http://www.isca-speech.org/archive/Interspeech_2016/abstracts/0211.html}
\BIBentrySTDinterwordspacing

\bibitem{Eyben2013Real-life}
F.~Eyben, F.~Weninger, S.~Squartini, and B.~Schuller, ``Real-life voice
  activity detection with lstm recurrent neural networks and an application to
  hollywood movies.''\hskip 1em plus 0.5em minus 0.4em\relax 2013 IEEE
  International Conference on Acoustics, Speech and Signal Processing, 5 2013,
  pp. 483--487.

\bibitem{Weninger2015Speech}
\BIBentryALTinterwordspacing
F.~Weninger, H.~Erdogan, S.~Watanabe, E.~Vincent, J.~Le~Roux, J.~R. Hershey,
  and B.~Schuller, ``\BIBforeignlanguage{en}{Speech enhancement with lstm
  recurrent neural networks and its application to noise-robust asr},'' in
  \emph{\BIBforeignlanguage{en}{Latent Variable Analysis and Signal
  Separation}}, E.~Vincent, A.~Yeredor, Z.~Koldovský, and P.~Tichavský,
  Eds.\hskip 1em plus 0.5em minus 0.4em\relax Cham: Springer International
  Publishing, 2015, vol. 9237, pp. 91--99, dOI: 10.1007/978-3-319-22482-4_11.
  [Online]. Available:
  \url{http://link.springer.com/10.1007/978-3-319-22482-4_11}
\BIBentrySTDinterwordspacing

\bibitem{Chen2015Speech}
Z.~Chen, S.~Watanabe, H.~Erdogan, and J.~R. Hershey, ``Speech enhancement and
  recognition using multi-task learning of long short-term memory recurrent
  neural networks,'' 2015.

\bibitem{Sun2017Multiple-target}
L.~Sun, J.~Du, L.~Dai, and C.~Lee, ``Multiple-target deep learning for lstm-rnn
  based speech enhancement.''\hskip 1em plus 0.5em minus 0.4em\relax 2017
  Hands-free Speech Communications and Microphone Arrays (HSCMA), 3 2017, pp.
  136--140.

\bibitem{Bitzer1999Multi-Microphone}
J.~Bitzer, K.~U. Simmer, and K.-d. Kammeyer, \emph{Multi-Microphone Noise
  Reduction Techniques For Hands-Free Speech Recognition - A Comparative
  Study}, 1999.

\bibitem{Liu1995Room}
Q.~Liu, B.~Champagne, and P.~Kabal, ``Room speech dereverberation via
  minimum-phase and all-pass component processing of multi-microphone
  signals.''\hskip 1em plus 0.5em minus 0.4em\relax IEEE Pacific Rim Conference
  on Communications, Computers, and Signal Processing. Proceedings, 5 1995, pp.
  571--574.

\bibitem{Hoshuyama1999robust}
O.~Hoshuyama, A.~Sugiyama, and A.~Hirano, ``A robust adaptive beamformer for
  microphone arrays with a blocking matrix using constrained adaptive
  filters,'' \emph{IEEE Transactions on Signal Processing}, vol.~47, no.~10,
  pp. 2677--2684, 10 1999.

\bibitem{Yousefian2011Dual-Microphone}
\BIBentryALTinterwordspacing
N.~Yousefian and P.~Loizou, ``\BIBforeignlanguage{en}{A dual-microphone speech
  enhancement algorithm based on the coherence function},''
  \emph{\BIBforeignlanguage{en}{IEEE Transactions on Audio, Speech, and
  Language Processing}}, 2011, [Online; accessed 2018-10-19]. [Online].
  Available: \url{http://ieeexplore.ieee.org/document/5957265/}
\BIBentrySTDinterwordspacing

\bibitem{Kailath1985Adaptive}
Kailath and T., ``Adaptive beamforming for coherent signals and interference,''
  \emph{IEEE Transactions on Acoustics, Speech, and Signal Processing},
  vol.~33, no.~3, pp. 527--536, 6 1985.

\bibitem{Li2018Distributed-microphones}
X.~Li, M.~Fan, L.~Liu, and W.~Li, ``Distributed-microphones based in-vehicle
  speech enhancement via sparse and low-rank spectrogram decomposition,''
  \emph{Speech Communication}, vol.~98, pp. 51--62, 4 2018.

\bibitem{Vincent2008Extracting}
\BIBentryALTinterwordspacing
P.~Vincent, H.~Larochelle, Y.~Bengio, and P.-A. Manzagol,
  ``\BIBforeignlanguage{en}{Extracting and composing robust features with
  denoising autoencoders},'' the 25th international conference.\hskip 1em plus
  0.5em minus 0.4em\relax Helsinki, Finland: ACM Press, 2008, pp. 1096--1103,
  [Online; accessed 2018-10-26]. [Online]. Available:
  \url{http://portal.acm.org/citation.cfm?doid=1390156.1390294}
\BIBentrySTDinterwordspacing

\bibitem{Araki2015Exploring}
S.~Araki, T.~Hayashi, M.~Delcroix, M.~Fujimoto, K.~Takeda, and T.~Nakatani,
  ``Exploring multi-channel features for denoising-autoencoder-based speech
  enhancement.''\hskip 1em plus 0.5em minus 0.4em\relax 2015 IEEE International
  Conference on Acoustics, Speech and Signal Processing (ICASSP), 4 2015, pp.
  116--120.

\bibitem{Wang2018All-Neural}
\BIBentryALTinterwordspacing
Z.-Q. Wang and D.~Wang, ``\BIBforeignlanguage{en}{All-neural multi-channel
  speech enhancement},'' Interspeech 2018.\hskip 1em plus 0.5em minus
  0.4em\relax ISCA, 9 2018, pp. 3234--3238, [Online; accessed 2018-10-26].
  [Online]. Available:
  \url{http://www.isca-speech.org/archive/Interspeech_2018/abstracts/1664.html}
\BIBentrySTDinterwordspacing

\bibitem{Fu2017Raw}
S.-W. Fu, Y.~Tsao, X.~Lu, and H.~Kawai, ``Raw waveform-based speech enhancement
  by fully convolutional networks,'' \emph{Proceedings - 9th Asia-Pacific
  Signal and Information Processing Association Annual Summit and Conference,
  APSIPA ASC 2017}, vol. 2018-Febru, pp. 6--12, 2017.

\bibitem{Fu2018End-to-End}
S.-W. Fu, T.-W. Wang, Y.~Tsao, X.~Lu, and H.~Kawai,
  ``\BIBforeignlanguage{en}{End-to-end waveform utterance enhancement for
  direct evaluation metrics optimization by fully convolutional neural
  networks},'' \emph{\BIBforeignlanguage{en}{IEEE/ACM Transactions on Audio,
  Speech, and Language Processing}}, vol.~26, no.~9, pp. 1570--1584, 9 2018.

\bibitem{Pascual2017SEGAN:}
S.~Pascual, A.~Bonafonte, and J.~Serra, ``Segan: Speech enhancement generative
  adversarial network,'' vol. 2017-Augus, no.~D, 2017, pp. 3642--3646.

\bibitem{Rethage2017Wavenet}
D.~Rethage, J.~Pons, and X.~Serra, ``A wavenet for speech denoising,'' 2017.

\bibitem{Qian2017Speech}
\BIBentryALTinterwordspacing
K.~Qian, Y.~Zhang, S.~Chang, X.~Yang, D.~Florêncio, and M.~Hasegawa-Johnson,
  ``\BIBforeignlanguage{en}{Speech enhancement using bayesian wavenet},''
  Interspeech 2017.\hskip 1em plus 0.5em minus 0.4em\relax ISCA, 8 2017, pp.
  2013--2017, [Online; accessed 2019-04-24]. [Online]. Available:
  \url{http://www.isca-speech.org/archive/Interspeech_2017/abstracts/1672.html}
\BIBentrySTDinterwordspacing

\bibitem{Grais2017Multi-Resolution}
\BIBentryALTinterwordspacing
E.~M. Grais, H.~Wierstorf, D.~Ward, and M.~D. Plumbley, ``Multi-resolution
  fully convolutional neural networks for monaural audio source separation,''
  \emph{arXiv:1710.11473 [cs, eess]}, 10 2017, arXiv: 1710.11473. [Online].
  Available: \url{http://arxiv.org/abs/1710.11473}
\BIBentrySTDinterwordspacing

\bibitem{Grais2018Raw}
\BIBentryALTinterwordspacing
E.~M. Grais, D.~Ward, and M.~D. Plumbley, ``Raw multi-channel audio source
  separation using multi-resolution convolutional auto-encoders,''
  \emph{arXiv:1803.00702 [cs]}, 3 2018, arXiv: 1803.00702. [Online]. Available:
  \url{http://arxiv.org/abs/1803.00702}
\BIBentrySTDinterwordspacing

\bibitem{Taal2010short-time}
\BIBentryALTinterwordspacing
C.~H. Taal, R.~C. Hendriks, R.~Heusdens, and J.~Jensen,
  ``\BIBforeignlanguage{en}{A short-time objective intelligibility measure for
  time-frequency weighted noisy speech}.''\hskip 1em plus 0.5em minus
  0.4em\relax IEEE, 2010, pp. 4214--4217, [Online; accessed 2018-07-31].
  [Online]. Available: \url{http://ieeexplore.ieee.org/document/5495701/}
\BIBentrySTDinterwordspacing

\bibitem{Taal2011Algorithm}
------, ``\BIBforeignlanguage{en}{An algorithm for intelligibility prediction
  of time–frequency weighted noisy speech},''
  \emph{\BIBforeignlanguage{en}{IEEE Transactions on Audio, Speech, and
  Language Processing}}, vol.~19, no.~7, pp. 2125--2136, 9 2011.

\bibitem{RixPerceptual}
A.~W. Rix, J.~G. Beerends, M.~P. Hollier, and A.~P. Hekstra,
  ``\BIBforeignlanguage{en}{Perceptual evaluation of speech quality (pesq)-a
  new method for speech quality assessment of telephone networks and codecs},''
  p.~4.

\bibitem{Oord2016WaveNet:}
\BIBentryALTinterwordspacing
v.~d.~A. Oord, S.~Dieleman, H.~Zen, K.~Simonyan, O.~Vinyals, A.~Graves,
  N.~Kalchbrenner, A.~Senior, and K.~Kavukcuoglu,
  ``\BIBforeignlanguage{en}{Wavenet: A generative model for raw audio},''
  \emph{\BIBforeignlanguage{en}{arXiv:1609.03499 [cs]}}, 9 2016, arXiv:
  1609.03499. [Online]. Available: \url{http://arxiv.org/abs/1609.03499}
\BIBentrySTDinterwordspacing

\bibitem{Ravanelli2018Speaker}
\BIBentryALTinterwordspacing
M.~Ravanelli and Y.~Bengio, ``\BIBforeignlanguage{en}{Speaker recognition from
  raw waveform with sincnet},'' \emph{\BIBforeignlanguage{en}{arXiv:1808.00158
  [cs, eess]}}, 7 2018, arXiv: 1808.00158. [Online]. Available:
  \url{http://arxiv.org/abs/1808.00158}
\BIBentrySTDinterwordspacing

\bibitem{Luo2018TasNet:}
\BIBentryALTinterwordspacing
Y.~Luo and N.~Mesgarani, ``Tasnet: Surpassing ideal time-frequency masking for
  speech separation,'' \emph{arXiv:1809.07454 [cs, eess]}, 9 2018, arXiv:
  1809.07454. [Online]. Available: \url{http://arxiv.org/abs/1809.07454}
\BIBentrySTDinterwordspacing

\bibitem{Donahue2018Adversarial}
\BIBentryALTinterwordspacing
C.~Donahue, J.~McAuley, and M.~Puckette, ``Adversarial audio synthesis,''
  \emph{arXiv:1802.04208 [cs]}, 2 2018, arXiv: 1802.04208. [Online]. Available:
  \url{http://arxiv.org/abs/1802.04208}
\BIBentrySTDinterwordspacing

\bibitem{Yu2015Multi-Scale}
\BIBentryALTinterwordspacing
F.~Yu and V.~Koltun, ``Multi-scale context aggregation by dilated
  convolutions,'' \emph{arXiv:1511.07122 [cs]}, 11 2015, arXiv: 1511.07122.
  [Online]. Available: \url{http://arxiv.org/abs/1511.07122}
\BIBentrySTDinterwordspacing

\bibitem{Zhang2017Speech}
\BIBentryALTinterwordspacing
A.~Zhang, \emph{Speech Recognition (Version 3.8) [Software]. Available from
  https://github.com/Uberi/speech_recognition.}, 2017, original-date:
  2014-04-23T04:53:54Z. [Online]. Available:
  \url{https://github.com/Uberi/speech_recognition}
\BIBentrySTDinterwordspacing

\bibitem{Lai2018Deep}
Y.-H. Lai, Y.~Tsao, X.~Lu, F.~Chen, Y.-T. Su, K.-C. Chen, Y.-H. Chen, L.-C.
  Chen, L.~Po-Hung~Li, and C.-H. Lee, ``\BIBforeignlanguage{en}{Deep
  learning–based noise reduction approach to improve speech intelligibility
  for cochlear implant recipients:},'' \emph{\BIBforeignlanguage{en}{Ear and
  Hearing}}, vol.~39, no.~4, pp. 795--809, 2018.

\bibitem{Lee2018Speech}
W.-J. Lee, S.-S. Wang, F.~Chen, X.~Lu, S.-Y. Chien, and Y.~Tsao, ``Speech
  dereverberation based on integrated deep and ensemble learning algorithm,''
  \emph{2018 IEEE International Conference on Acoustics, Speech and Signal
  Processing (ICASSP)}, pp. 5454--5458, 2018.

\bibitem{Liu2018Bone-conducted}
H.-P. Liu, Y.~Tsao, and C.-S. Fuh, ``\BIBforeignlanguage{en}{Bone-conducted
  speech enhancement using deep denoising autoencoder},''
  \emph{\BIBforeignlanguage{en}{Speech Communication}}, vol. 104, pp. 106--112,
  11 2018.

\bibitem{Kingma2014Adam:}
\BIBentryALTinterwordspacing
D.~P. Kingma and J.~Ba, ``Adam: A method for stochastic optimization,''
  \emph{arXiv:1412.6980 [cs]}, 12 2014, arXiv: 1412.6980. [Online]. Available:
  \url{http://arxiv.org/abs/1412.6980}
\BIBentrySTDinterwordspacing

\bibitem{Kondo2006On}
K.~Kondo, T.~Fujita, and K.~Nakagawa, ``On equalization of bone conducted
  speech for improved speech quality.''\hskip 1em plus 0.5em minus 0.4em\relax
  2006 IEEE International Symposium on Signal Processing and Information
  Technology, 8 2006, pp. 426--431.

\bibitem{Bouserhal2017In-ear}
R.~E. Bouserhal, T.~H. Falk, and J.~Voix, ``In-ear microphone speech quality
  enhancement via adaptive filtering and artificial bandwidth extension,''
  \emph{The Journal of the Acoustical Society of America}, vol. 141, no.~3, pp.
  1321--1331, 3 2017.

\bibitem{Wong2007Development}
L.~L.~N. Wong, S.~D. Soli, S.~Liu, N.~Han, and M.-W. Huang,
  ``\BIBforeignlanguage{en}{Development of the mandarin hearing in noise test
  (mhint):},'' \emph{\BIBforeignlanguage{en}{Ear and Hearing}}, vol.~28, no.
  Supplement, pp. 70S--74S, 4 2007.

\bibitem{PGA181}
\BIBentryALTinterwordspacing
``\BIBforeignlanguage{en-US}{Pga181 - side-address cardioid condenser
  microphone},'' [Online; accessed 2019-04-29]. [Online]. Available:
  \url{https://www.shure.com/en-US/products/microphones/pga181}
\BIBentrySTDinterwordspacing

\bibitem{SANLUX}
\BIBentryALTinterwordspacing
``Sanlux hmt-11,'' [Online; accessed 2019-04-29]. [Online]. Available:
  \url{http://www.sanyo.com.tw/s1504/sanyo_in_b.asp?model=2033}
\BIBentrySTDinterwordspacing

\bibitem{barker2015third}
J.~Barker, R.~Marxer, E.~Vincent, and S.~Watanabe, ``The third
  ‘chime’speech separation and recognition challenge: Dataset, task and
  baselines,'' in \emph{2015 IEEE Workshop on Automatic Speech Recognition and
  Understanding (ASRU)}.\hskip 1em plus 0.5em minus 0.4em\relax IEEE, 2015, pp.
  504--511.
  
\bibitem{garofalo2007csr}
J.~Garofalo, D.~Graff, D.~Paul, and D.~Pallett, ``Csr-i (wsj0) complete,''
  \emph{Linguistic Data Consortium, Philadelphia}, 2007.  

\bibitem{koizumi2018dnn}
Y.~Koizumi, K.~Niwa, Y.~Hioka, K.~Kobayashi, and Y.~Haneda, ``Dnn-based source
  enhancement to increase objective sound quality assessment score,''
  \emph{IEEE/ACM Transactions on Audio, Speech, and Language Processing},
  vol.~26, no.~10, pp. 1780--1792, 2018.
  
\bibitem{mittermaier2019small}
S.~Mittermaier, L.~K{\"u}rzinger, B.~Waschneck, and G.~Rigoll,
  ``Small-footprint keyword spotting on raw audio data with
  sinc-convolutions,'' \emph{arXiv preprint arXiv:1911.02086}, 2019.

\bibitem{gong2019dilated}
S.~Gong, Z.~Wang, T.~Sun, Y.~Zhang, C.~D. Smith, L.~Xu, and J.~Liu, ``Dilated
  fcn: Listening longer to hear better,'' in \emph{2019 IEEE Workshop on
  Applications of Signal Processing to Audio and Acoustics (WASPAA)}.\hskip 1em
  plus 0.5em minus 0.4em\relax IEEE, 2019, pp. 254--258.

\bibitem{paliwal2011importance}
K.~Paliwal, K.~W{\'o}jcicki, and B.~Shannon, ``The importance of phase in
  speech enhancement,'' \emph{speech communication}, vol.~53, no.~4, pp.
  465--494, 2011.

\bibitem{le2011phase}
J.~Le~Roux, ``Phase-controlled sound transfer based on maximally-inconsistent
  spectrograms,'' \emph{Signal}, vol.~5, p.~10, 2011.

\bibitem{gerkmann2015phase}
T.~Gerkmann, M.~Krawczyk-Becker, and J.~Le~Roux, ``Phase processing for
  single-channel speech enhancement: History and recent advances,'' \emph{IEEE
  Signal Processing Magazine}, vol.~32, no.~2, pp. 55--66, 2015.

\end{thebibliography}
\end{document}